\documentclass[prb,aps,superscriptaddress,twocolumn]{revtex4}
\usepackage{rotating}
\usepackage{graphicx}% Include figure files
\usepackage{dcolumn}% Align table columns on decimal point
\usepackage{bm}% bold math
\usepackage{amssymb}
\begin{document}

\title{Finite temperature phase diagram of the classical Kitaev-Heisenberg model}

\author{Craig Price}

\affiliation{Department of Physics, The Pennsylvania State University, 104 Davey Lab, University Park, Pennsylvania 16802, USA}

\author{Natalia B. Perkins}

\affiliation{Department of Physics, University of Wisconsin, 1150
University Ave., Madison, Wisconsin 53706, USA}

\begin{abstract}

We investigate the  finite-temperature phase diagram of the classical Kitaev-Heisenberg model on the hexagonal lattice.
Due to the anisotropy introduced  by the Kitaev interaction, the model is magnetically ordered at low temperatures.
The ordered phase is stabilized entropically by an order by disorder mechanism where thermal fluctuations of classical spins select collinear magnetic states in which magnetic moments point along one of the cubic directions.
We find that there is an intermediate phase between the low-temperature ordered phase and the high-temperature disordered phase.
We show that the intermediate phase  is  a critical Kosterlitz-Thouless phase exhibiting correlations of the order parameter that decay algebraically in space.
Using finite size scaling analysis, we  determine the boundaries of the critical phase with reasonable accuracy.
We  show that the Kitaev interaction plays a crucial role in understanding the finite temperature properties of A$_2$IrO$_3$ systems.

\end{abstract}
\date{\today}
\maketitle
\section{Introduction}

Recently  spin-orbit coupling (SOC) effects have become  a subject of intense research across many different disciplines in condensed matter physics.
These effects are  especially pronounced in 4d and 5d transition-metal compounds whose significant atomic SOC is due to their large atomic weight.

The strongly entangled, spin-orbital nature of the localized states are characterized by an effective angular momenta, $J_{\rm eff}$.
The interactions among the components of  $J_{\rm eff}$ are determined by the combination of spin and lattice symmetries.
This combination creates unconventional anisotropic exchange interactions that give rise to various novel properties.~\cite{jackeli09,jackeli10,chern09}
For example, breaking the spin rotation symmetry permits magnetic Hamiltonians that contain  terms that are products of different components of spin operators.
Such terms, not allowed in the  traditionally studied SU(2) symmetric models, introduce a new source of frustration~\cite{balents10} and might drive the system towards quantum  spin liquid states.

A prominent model exemplifying these kinds of highly anisotropic interactions is the Kitaev model  on the honeycomb lattice.~\cite{kitaev06,nussinov13}
 The ground state of the Kitaev model is known exactly; it  is a spin liquid characterized by anyonic excitations with exotic fractional  statistics.
  Recently, Jackeli and Khaliullin\cite{jackeli09} suggested  that the Kitaev model could actually be realized within the honeycomb iridates with general formula A$_2$IrO$_3$,  since the anisotropic part of the interactions among Ir-ions  has the same form as  the Kitaev coupling.
   This suggestion has triggered a lot of  experimental~\cite{singh10,liu11,singh12,ye12,choi12} and theoretical~\cite{jiang11,reuther11,trousselet11,hyart12,price12,mazin12,ybkim12} activity in the study of Na$_2$IrO$_3$ and Li$_2$IrO$_3$ compounds.

In A$_2$IrO$_3$ materials, Ir$^{4+}$ ions  are in a  low spin $5d^5$ configuration with an effective angular momentum of $J_{\rm eff}=1/2$ due to strong SOC.
The low-energy Hamiltonian describing the interaction  between $J_{\rm eff}$ iridium moments is called the Kitaev-Heisenberg (KH) model since it contains both the  isotropic antiferromagnetic (AF) Heisenberg interaction, $J_H$, and the anisotropic ferromagnetic (FM) Kitaev interaction, $J_K$.
 The Kitaev  exchange interaction is generated through the 90$^\circ$ Ir-O-Ir hopping path between $J_{\rm eff}$ Kramers doublets and  is non-zero only  in the presence of  finite  Hund's coupling.
 The isotropic exchange via the 90$^\circ$ Ir-O-Ir  is canceled due to a destructive interference among multiple superexchange paths.
At the same time, the isotropic AF Heisenberg interaction is suppressed and the only non-vanishing contribution is from the superexchange interaction arising  from the direct overlap of the Ir 5d orbitals.

The ground state of the KH model can be a spin liquid  despite the presence of the direct isotropic exchange.
 This is because the Kitaev spin liquid is rather stable with respect to the Heisenberg interaction \cite{jackeli10,jiang11,trousselet11,ybkim12}; it remains the ground state of the KH model for a wide range of strengths of the Heisenberg  interaction.
 Exact diagonalization studies \cite{jackeli10,jiang11,trousselet11} suggest a stability of the spin liquid phase for the model parameter $\alpha$ in the range $(0.8, 1)$, where  $\alpha$ is determined  such that $J_K=2\alpha$ and $J_H=1-\alpha$.
 Even so, the Kitaev  spin liquid in honeycomb iridates  has not been  observed yet;  in both Na$_2$IrO$_3$ and Li$_2$IrO$_3$,  magnetic order has been observed at low temperatures.~\cite{singh10,liu11,singh12}
 Moreover for Na$_2$IrO$_3$, the KH model in its original form does not appear to be sufficient to account for neither the  zigzag magnetic order nor the spectrum of magnetic excitations that have been measured in neutron scattering experiments.~\cite{liu11,singh12,choi12}

 To explain the  experimental observations in Na$_2$IrO$_3$, three different modifications of the super-exchange model  have been  proposed.~\cite{kimchi11,chaloupka12,Subhro12}
 In the first approach,~\cite{kimchi11} it was shown that the zigzag magnetic order may be stabilized within the KH model by including substantial second and third neighbour antiferromagnetic interactions.
The second approach~\cite{chaloupka12}  extends the KH model to its full parameter space by including additional hopping processes based on the $t_{2g}-e_g$ hopping  along the 90$^\circ$ Ir-O-Ir paths.
The main difference between these two approaches is that  the role of the Kitaev interaction is  minor in the first approach while it still plays the dominant role in the second one.
 The third approach assumes that Na$_2$IrO$_3$ is significantly distorted from the ideal structure; here, the trigonal distortion  is the dominant  interaction while the SOC is subdominant.
   This third model actually  does not contain the Kitaev term at all, but it does retain the zigzag ground state as one of its  magnetic ground states.~\cite{Subhro12}
   Thus, the importance of the Kitaev term in the  low-energy physics  of the A$_2$IrO$_3$ compounds remains an open question.

\begin{figure}
\includegraphics[width=0.95\columnwidth]{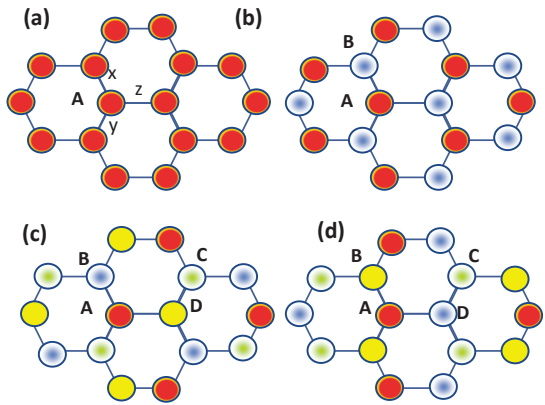}
\caption{
Four possible magnetic configurations:
(a) FM ordering;
(b) two-sublattice, AF N\'{e}el order;
(c) stripy  order;
(d) zigzag order.
Open and filled circles correspond to up and down spins.
}
\label{fig:orders}
\end{figure}

In this paper,  we   continue to examine the role of the Kitaev interaction in the magnetic properties of A$_2$IrO$_3$ systems  using the  KH model  defined in both the original~\cite{jackeli10} and the extended~\cite{chaloupka12} parameter space.
We argue that the Kitaev interaction's main effect is the reduction of the symmetry of the system  from the continuous SU(2) symmetry to a discrete $\mathbb{Z}_6$ symmetry.
The latter is crucial for the finite temperature properties of quasi two-dimensional (2D) iridates since by the Mermin-Wagner theorem~\cite{mermin66} the 2D magnetic systems with continuous symmetry do not exhibit long-range magnetic order at any finite temperature.
 However, there is no such constraint  on spin systems with discrete symmetry.
 Thus, we argue that  the  Kitaev term is responsible for the  presence of the long range magnetic order.

 We also show  that the finite temperature properties  of the KH model are similar to those of the six-state clock model.~\cite{jose77,ortiz12}
  The KH model undergoes two continuous phase transitions as a function of temperature.
 This gives rise to three different phases: a low-T ordered phase with a spontaneously broken $\mathbb{Z}_6$ symmetry, an intermediate  critical phase, and a high-T disordered phase.
Finite size scaling analysis  of our Monte  Carlo (MC) simulations confirmed that the intermediate phase is a critical Kosterlitz-Thouless (KT)  phase with floating exponents and algebraic correlations.

The rest of the paper is organized as follows.
We start our discussion with Sec.~\ref{model} which contains a brief summary of known facts about the classical KH model that will be used in the rest of the paper.
In Sec.~\ref{finite}, we discuss in detail  the results of our  numerical simulations.
  We analyze the  discreteness of the low-temperature phase, and then we  discuss the criticality of the intermediate phase and show how the finite-size scaling analysis allows us to  determine the boundary of the critical phase.
  We  show that the phase transitions are driven by the order by disorder mechanism, in which thermal fluctuations of classical spins select  collinear spin configurations where the magnetic moments all point along one of the cubic directions.
We also demonstrate that at particular points of the phase diagram, for which the continuous symmetry is preserved, the ordered state is destroyed at any non-zero temperature.
The  section ends with a discussion of  the finite temperature phase diagram of the classical KH model.

 In Sec.~\ref{PDextended}, we analyze the finite temperature properties of the KH model in its extended parameter space.
 We show that for the parameters relevant to the  Na$_2$IrO$_3$ and Li$_2$IrO$_3$ compounds, the model exhibits a zigzag magnetic order in agreement with experimental findings.
 In Sec.\ref{cubic},  we show that there is a significant difference between the KH model and the Heisenberg model with a cubic anisotropy.
  While the later model also exhibits two phase transitions,  its intermediate phase is not critical; it is nematic like.
    Using finite-size scaling analysis, we show that in this case the phase transitions are of the  three-states Potts  and Ising universality classes.
Sec.~\ref{conclusion} contains a summary of the obtained results and conclusions.

\begin{figure}
\includegraphics[width=0.95\columnwidth]{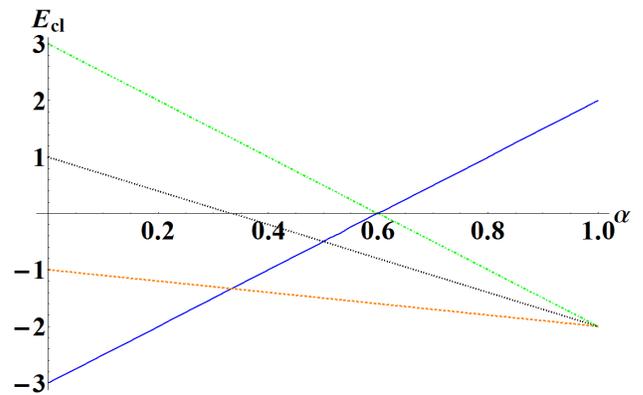}
\caption{
Classical energy as a function of $\alpha$: in the FM phase ($E_{cl}^{{M}}=3-5\alpha$, dot-dashed green line), in the zigzag phase ($E_{cl}^ {{Z}}=-3\alpha+1$, dotted black line), in the N\'{e}el phase ($E_{cl}^{{N}}=5\alpha-3$ solid blue line), and  in the stripy phase ($E_{cl}^{{S}}=-\alpha-1$ dashed red line).
}
\label{fig:cl-energy}
\end{figure}

\begin{figure*}
\includegraphics[width=0.49\columnwidth]{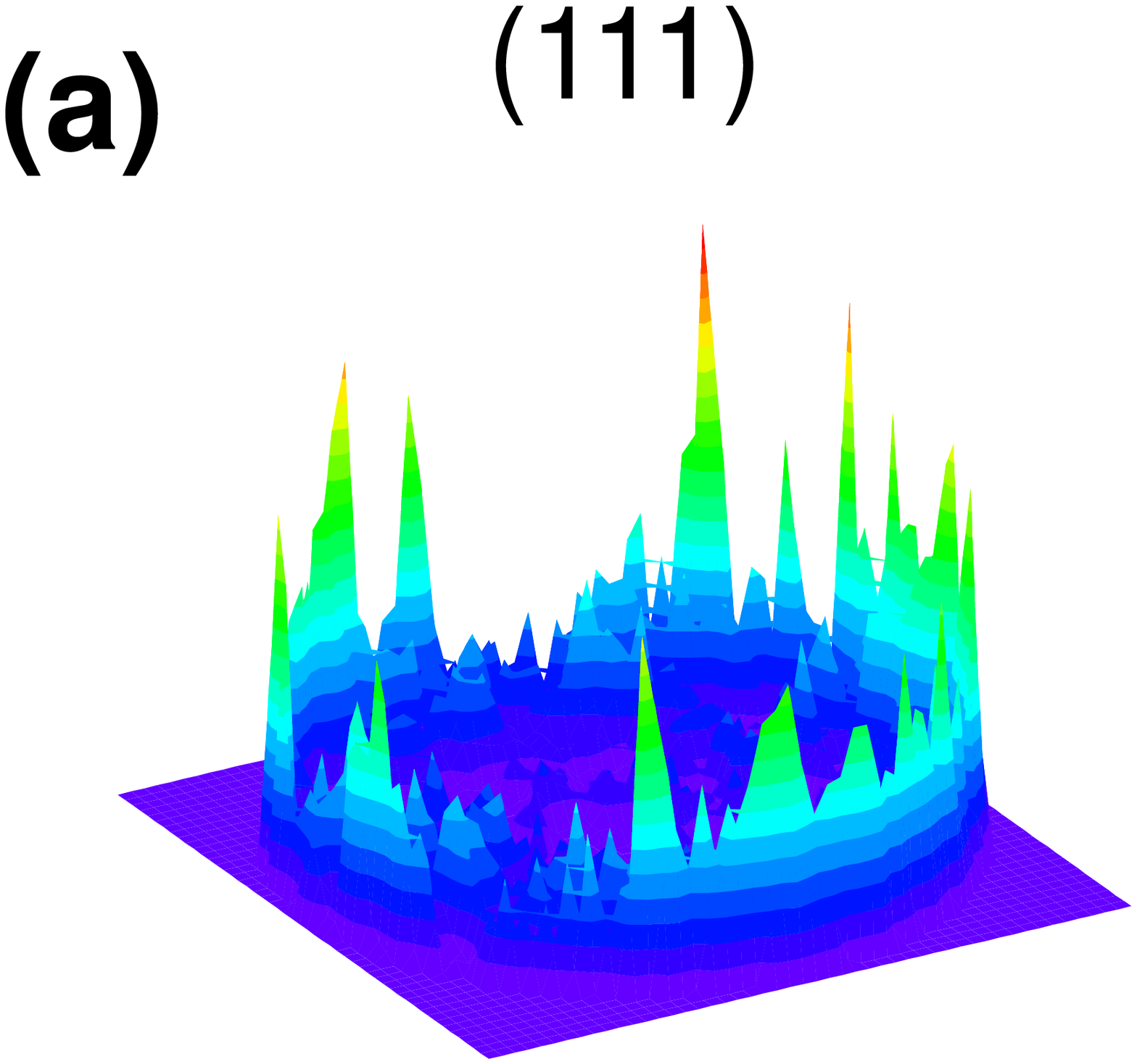}
\includegraphics[width=0.49\columnwidth]{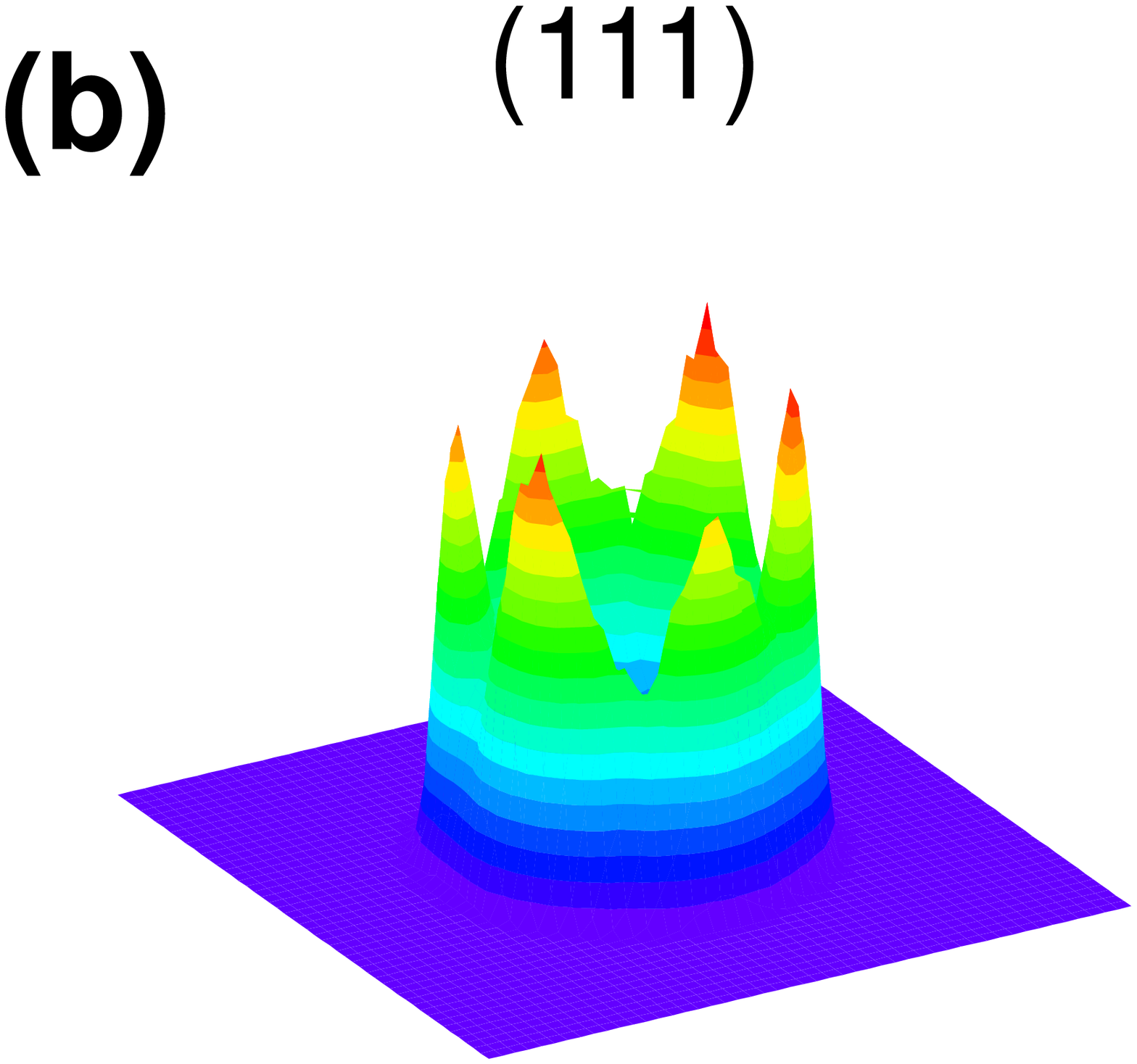}
\includegraphics[width=0.49\columnwidth]{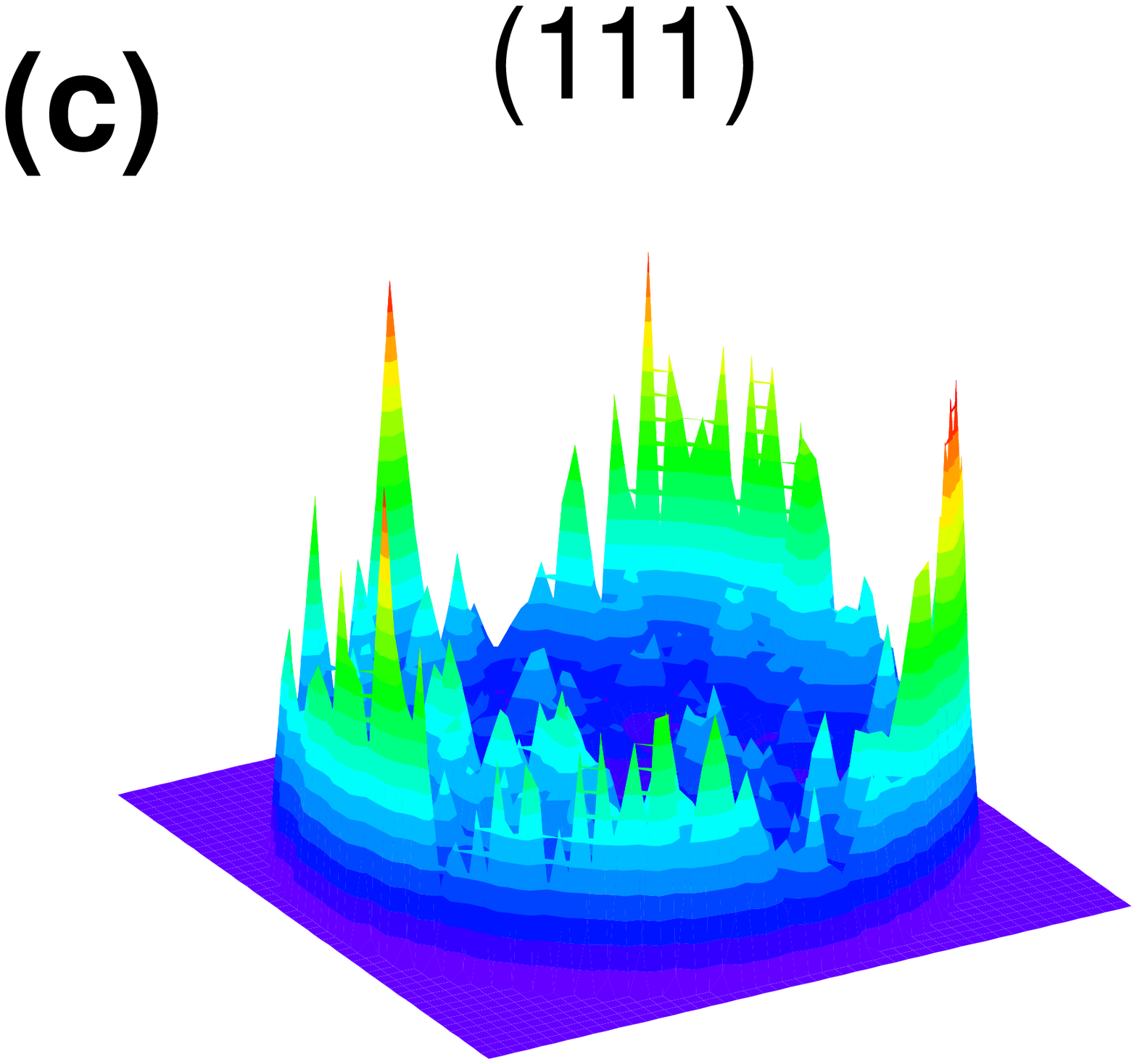}
\includegraphics[width=0.49\columnwidth]{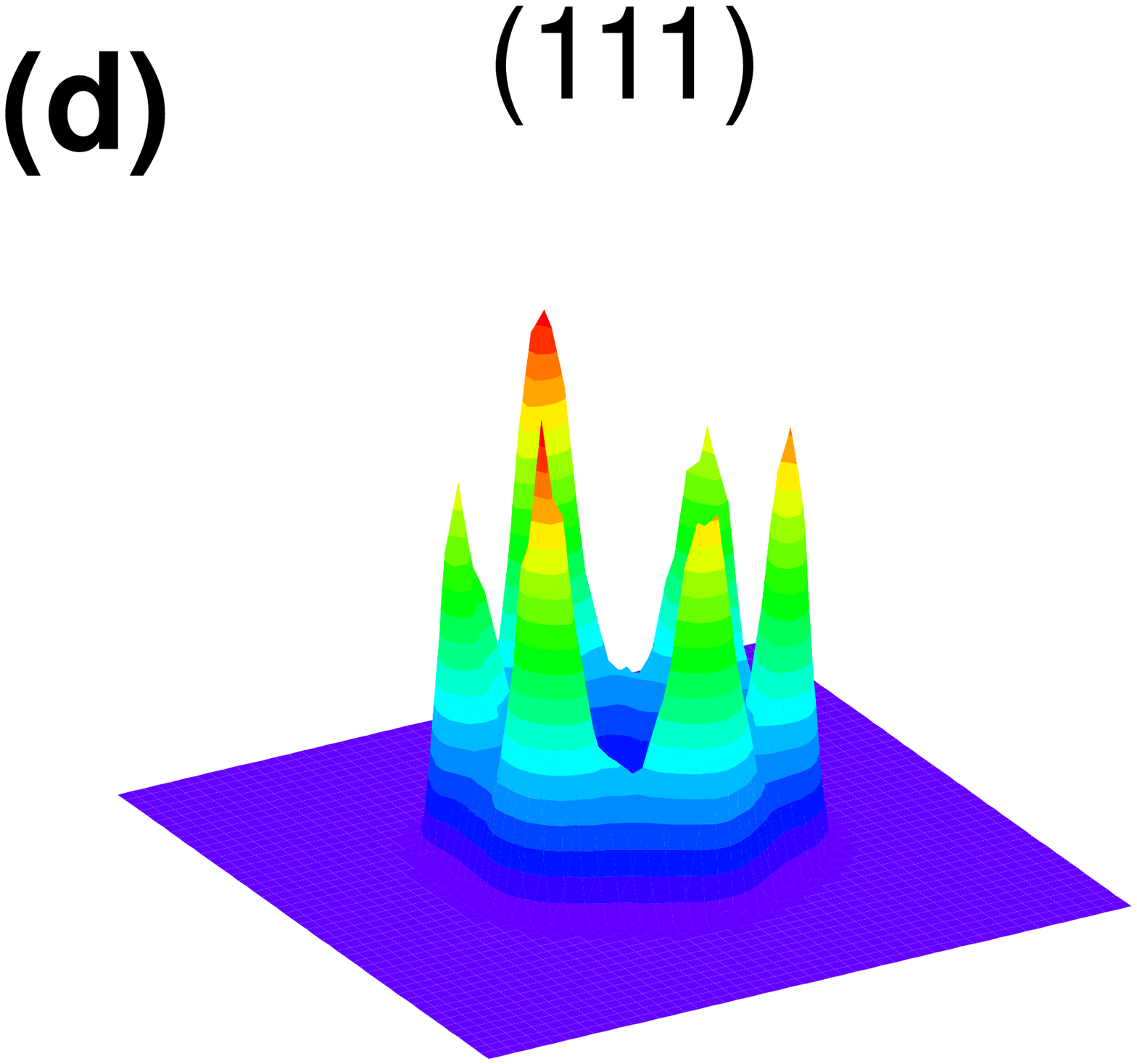}
\caption{
A low-T distribution of the projections of the vector order parameter on the (111) plane.
a) N\'{e}el order parameter  for $\alpha=0$;
 b) N\'{e}el order parameter  for $\alpha=0.25$;
c)  stripy order parameter for $\alpha=0.5$;
d) stripy order parameter for $\alpha=0.75$.
}
\label{fig:stereo}
\end{figure*}

\section{The classical KH Model}\label{model}

The classical version of the KH model is
\begin{eqnarray}
\label{ham1}
    \mathcal{H} =-J_K \sum_{\langle ij \rangle_\gamma} S_i^{\gamma}S_j^{\gamma}+ J_H\sum_{\langle ij \rangle} {\bf S}_i{\bf S}_j~.
\end{eqnarray}
Following  the notation of Ref. \onlinecite{jackeli10}, we denote the effective degrees of freedom of  Ir$^{4+}$ ions as  $S$, and take the quantization axes along the cubic axes of the IrO$_6$ octahedra.
$\gamma=x,y,z$ denotes the three bonds of the honeycomb lattice.
The two couplings, $J_H$ and $J_K$,  are opposite in sign;  $J_H$ is the AF isotropic exchange and $J_K$ is the FM anisotropic exchange.

The exchange constants  corresponding to  the  Kitaev and   the Heisenberg  interactions in the KH model (\ref{ham1}) can be conveniently described by one parameter, $\alpha$, such that $J_K=2\alpha$ and $J_H=1-\alpha$.  The model  then reads as
\begin{eqnarray}
\label{ham2}
    \mathcal{H} =-2\alpha \sum_{\langle ij \rangle_\gamma} S_i^{\gamma}S_j^{\gamma}+ (1-\alpha)\sum_{\langle ij \rangle} {\bf S}_i{\bf S}_j~.
\end{eqnarray}

\begin{figure*}
\includegraphics[width=0.55\columnwidth]{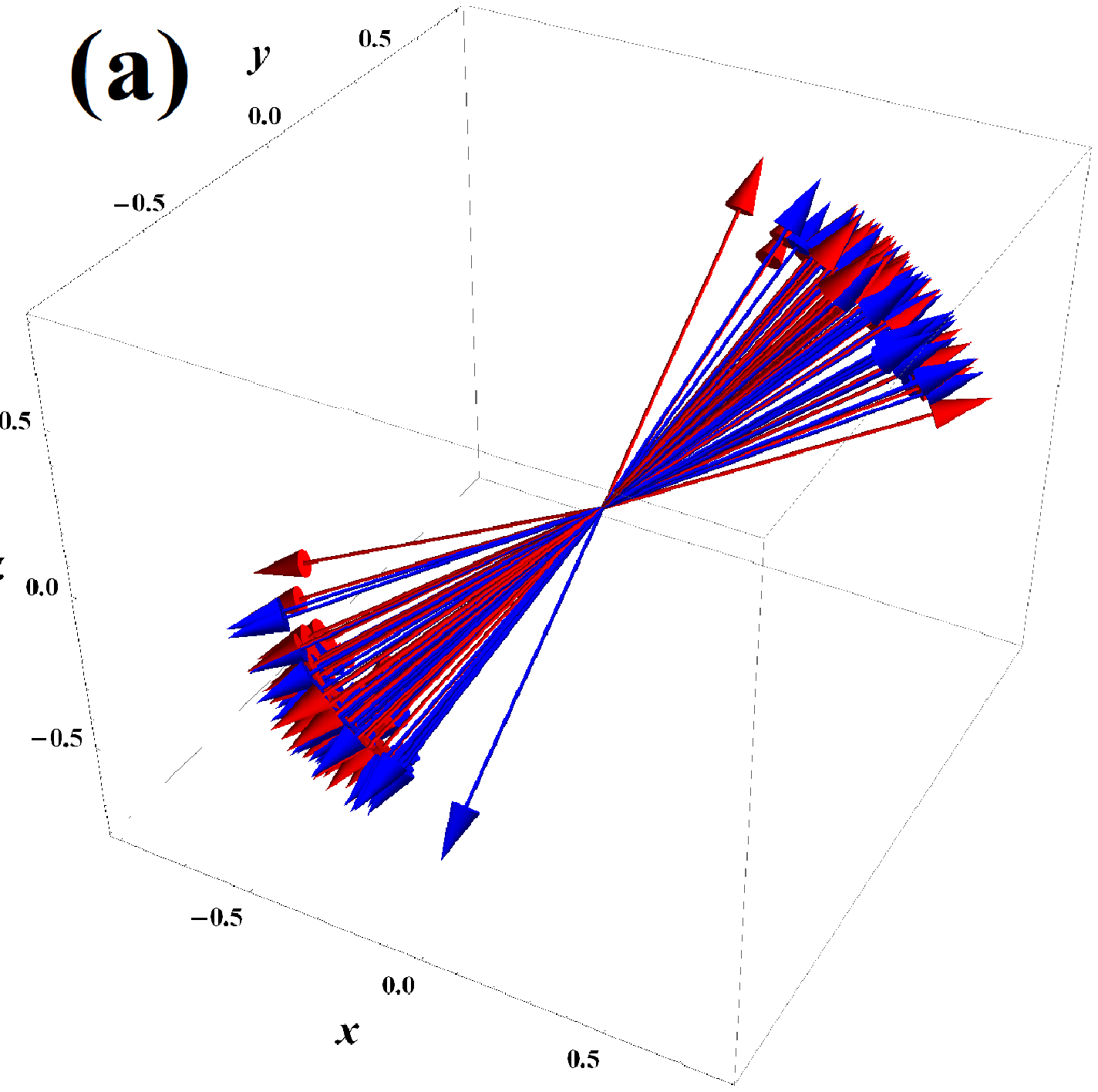}
\includegraphics[width=0.5\columnwidth]{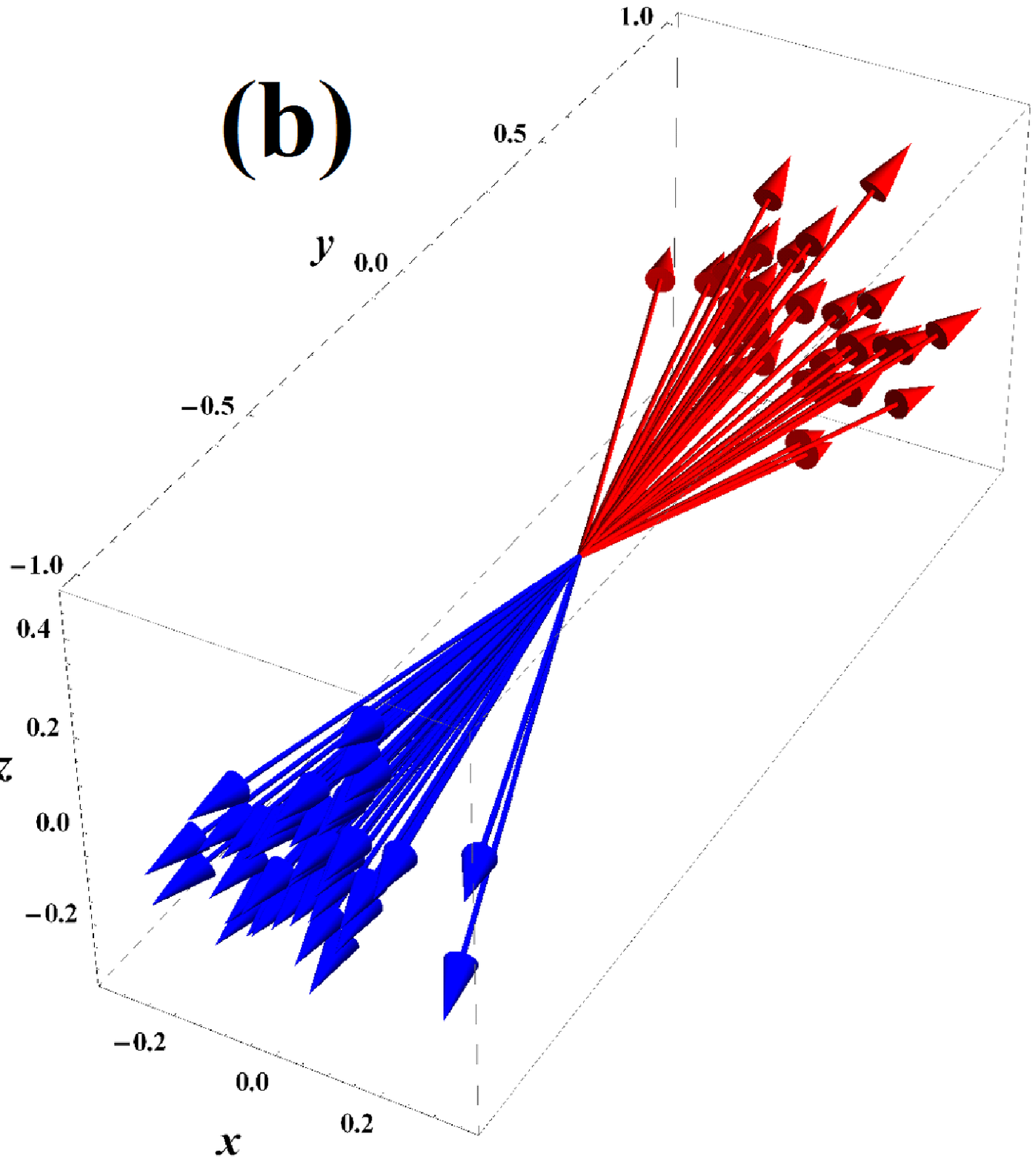}
\includegraphics[width=0.48\columnwidth]{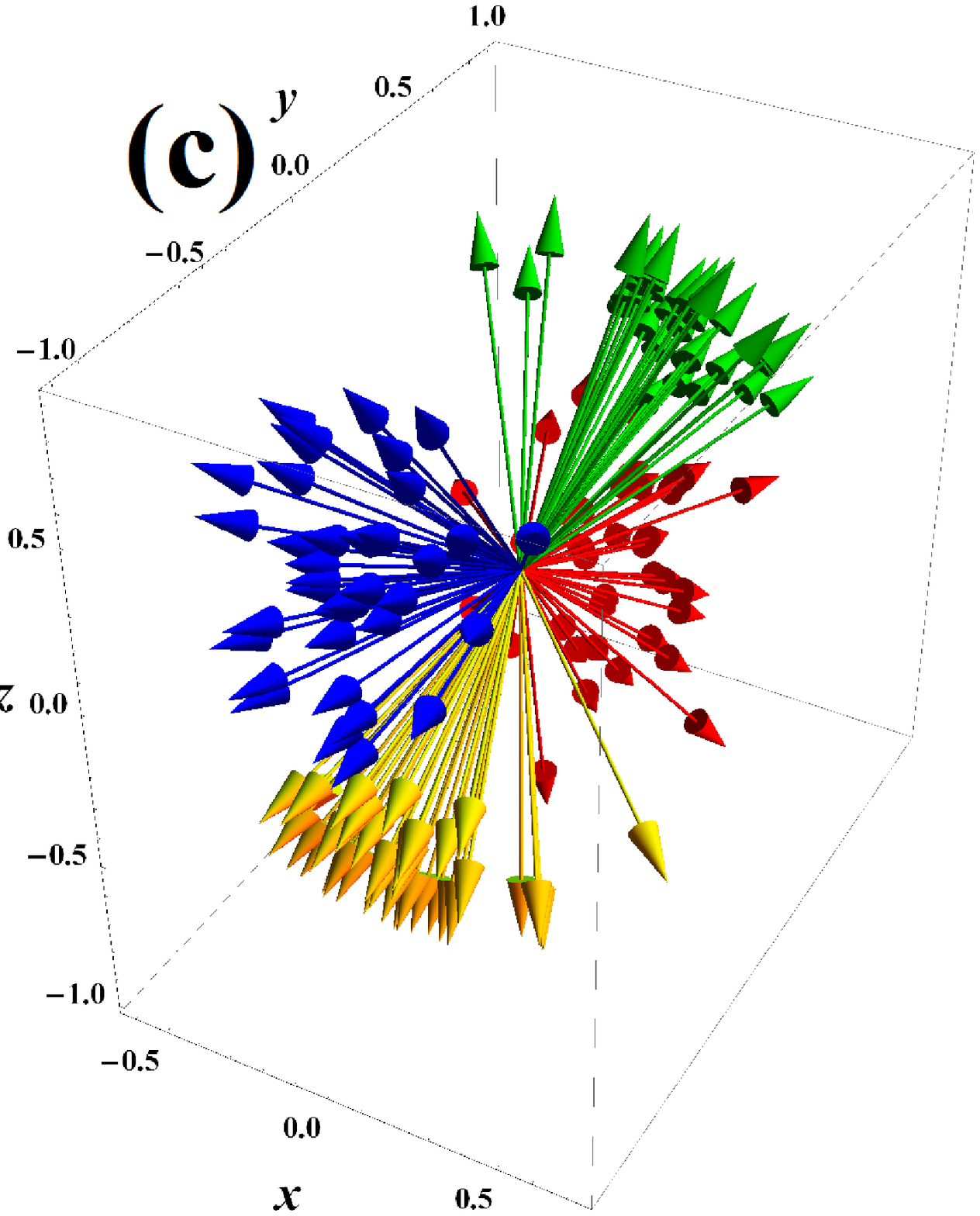}
\includegraphics[width=0.44\columnwidth]{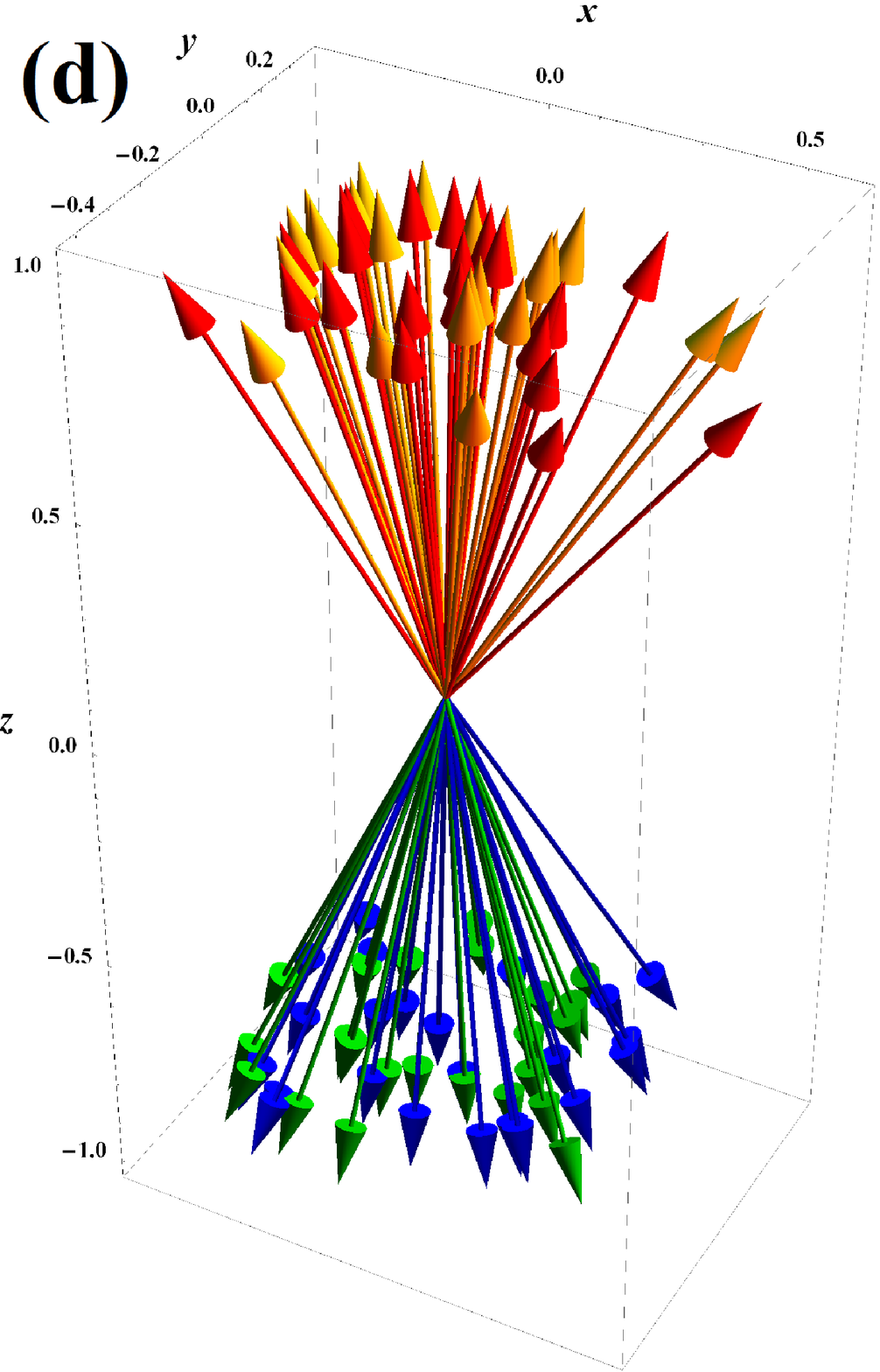}
\caption{Snapshots of spin configurations in the low-T phase  computed for a) $\alpha=0$, b) $\alpha=0.25$ c) $\alpha=0.5$ and d) $\alpha=0.75$. Each figure represents a single snapshot taken at a particular MC step.
Each color corresponds to one of the  two  or four different sublattices  describing the  N\'{e}el  or  the stripy order parameters, respectively.
}
\label{fig:snapshot}
\end{figure*}

 Let us briefly describe possible magnetically ordered states of model (\ref{ham2}).
 For this we need to introduce four sublattices: $A,B,C,$ and $D$.
For $0<\alpha<1/3$, the classical ground state is the simple two-sublattice AF N\'{e}el order (Fig.1 (b))  characterized by the order parameter
\begin{eqnarray}
\label{order:neel}
\mathbf{N}=\frac{1}{\mathcal{N}}\sum_{i}({\bf S}_{iA}-{\bf S}_{iB}).
\end{eqnarray}
${\mathcal{N}}$ denotes the number of sites.
The stripy  AF  order (Fig.1 (c)) describes the classical ground state of the model for   $1/3<\alpha<1$ and  is given by
\begin{eqnarray}
\label{order:stripy}
\mathbf{S}=\frac{1}{\mathcal{N}}\sum_{i}({\bf S}_{iA}-{\bf S}_{iB}-{\bf S}_{iC}+{\bf S}_{iD})~.
 \end{eqnarray}
 For $\alpha= 1.0$, the stripy  AF  state is classically degenerate with  other magnetically ordered states (see Fig.\ref{fig:cl-energy}).
 For example, it is degenerate with FM order (Fig.1 (a)) described by the total magnetization
\begin{eqnarray}
\label{order:ferro}
\mathbf{M}=\frac{1}{\mathcal{N}}\sum_{i}{\bf S}_{iA},
\end{eqnarray}
 and with zigzag AF spin order (Fig.1 (d)) described by
\begin{eqnarray}\label{order:zigzag}
  \mathbf{Z}=\frac{1}{\mathcal{N}}\sum_{i}({\bf S}_{iA}+{\bf S}_{iB}-{\bf S}_{iC}-{\bf S}_{iD})~.
 \end{eqnarray}

The classical degeneracy of the point corresponding to the classical Kitaev model is macroscopic and is known exactly; asymptotically, it has $(1.662)^N$ spin configurations.
This was computed by Baskaran {\it et al}~\cite{baskaran08}  by mapping  ordered states  of the classical Kitaev model to a certain dimer covering of the honeycomb lattice whose total number is known.\cite{baxter}

Because of the presence of the anisotropic Kitaev interaction term, the KH model has discrete spin-rotation symmetry for all of $\alpha$ except at two points, $\alpha=0$ and $\alpha=0.5$.
 At $\alpha=0$,   the KH model reduces to the antiferromagnetic Heisenberg model with  continuous SU(2) symmetry.
 At $\alpha=0.5$,  the stripy phase  becomes an exact ground state  of the model; it corresponds to a fully polarized FM state  in a rotated basis.
 This basis can be seen  by fixing the spin's direction on sublattice A  and  rotating the spins on sublattices $B, C,$ and $D$  by the angle $\pi$  about the $x,y$, and $z$ axis, respectively (see Fig.~\ref{fig:orders}(c)).~\cite{jackeli10}
   It is evident that the FM state has true SU(2) symmetry.
  Away from these special points,  the symmetry of the KH model is discrete; it combines the cubic symmetry of both the spin  and the lattice space.

\section{Finite temperature phase diagram  and critical properties of the classical KH model}\label{finite}

In this section we study the behavior of the classical KH model (\ref{ham2}) at finite temperature using MC simulations based on the standard Metropolis algorithm.
 In our simulations, we treat the spins as three-dimensional (3D) vectors, ${\bf S}_i=(S_i^x ,S_i^y ,S_i^z)$, of unit magnitude with $(S_i^x)^2 + (S_i^y)^2+ (S_i^z)^2=1$.
At each temperature, more than $10^7$ MC sweeps  were performed.
Of these, $10^6$ were used to equilibrate the system, and afterwards only 1 out of every 5 sweeps was used to calculate the averages of physical quantities.
We  present all energies in units of $J_H$ and assume  $k_B=1$.
The simulations were performed on  different  systems with a total number of sites equal to $2*L*L$.
The systems are spanned by the primitive vectors of a triangular lattice ${\bf a}_1=(1/2,\sqrt{3}/2)$ and ${\bf a}_2=(1,0)$ with a 2-point basis using periodic boundary conditions.

In our simulations we compute the following observables: four different order parameters ${\mathbf O}=\{{\mathbf N}, {\mathbf S}, {\mathbf M}, {\mathbf Z}\}$ defined in Eqs.(\ref{order:neel}-\ref{order:zigzag}), corresponding susceptibilities
\begin{eqnarray}\label{chi}
\chi_{\mathbf O}={\mathcal N}(\langle {\mathbf O}^2\rangle-\langle {\mathbf O}\rangle^2)/T~,
\end{eqnarray}
the Binder's cumulants
\begin{eqnarray}\label{binder}
B_{\mathbf O}=1-\langle {\mathbf O}^4\rangle/3\langle {\mathbf O}^2\rangle^2~,
\end{eqnarray}
 and the specific heat
\begin{eqnarray}\label{specific}
 C=(\langle E^2\rangle-\langle E\rangle^2)/{\mathcal N}T^2~.
 \end{eqnarray}

\subsection{Low-temperature ordered phase}

At low temperatures, the KH model  magnetically orders   in either a N\'{e}el state,  ${\mathbf N}$,  or  in a stripy state, ${\mathbf S}$, depending on the relative strengths of  the Kitaev and the Heisenberg interactions.
The presence of long range order at finite temperatures requires a discreteness of the  order parameter, which means that  the direction of the order parameter must also be selected.
This, however, does not happen on the level of non-interacting spin waves.
For both the N\'{e}el state and   the stripy state, the linear spin-wave spectrum has a quasi-Goldstone mode at the ordering vector  simulating the  spontaneous breaking of the continuous symmetry.\cite{jackeli10}
The discrete symmetry of the order parameter only appears  due to the contribution of higher order anharmonic modes of spin fluctuations.
These fluctuations lower the energy of the states whose order parameter points along  cubic directions~\cite{jackeli10} and thus removes  the accidental degeneracy  of the classical ground states.
This gives a  six-fold  degeneracy of the order parameter manifold which corresponds  to the six degenerate minima in free energy.
This is a well known  order by disorder mechanism in which  spin fluctuations (quantum or thermal) remove the accidental degeneracy and select the true ordered state.

The discreteness of the order parameter at all but  special points can be revealed from the histogram method in which  we record  two-dimensional distributions of the projections of the vector order parameter on the (111) plane.
We use the projection of the order parameter on the (111) plane  because  it  preserves  the cubic symmetry of the model and shows that its degeneracy  is related to the orientation of the order parameter with  respect to the directions in a cubic crystal.
In  Fig.~\ref{fig:stereo} (a), (b), (c), and (d) we present, respectively,   the low-T distributions of the  projection of the vector order parameter on the (111) plane  for  four different values of $\alpha$: $0,~\,~0.25,~\,0.5,$ and$~\,0.75$.
\begin{figure*}
\includegraphics[width=0.517\columnwidth]{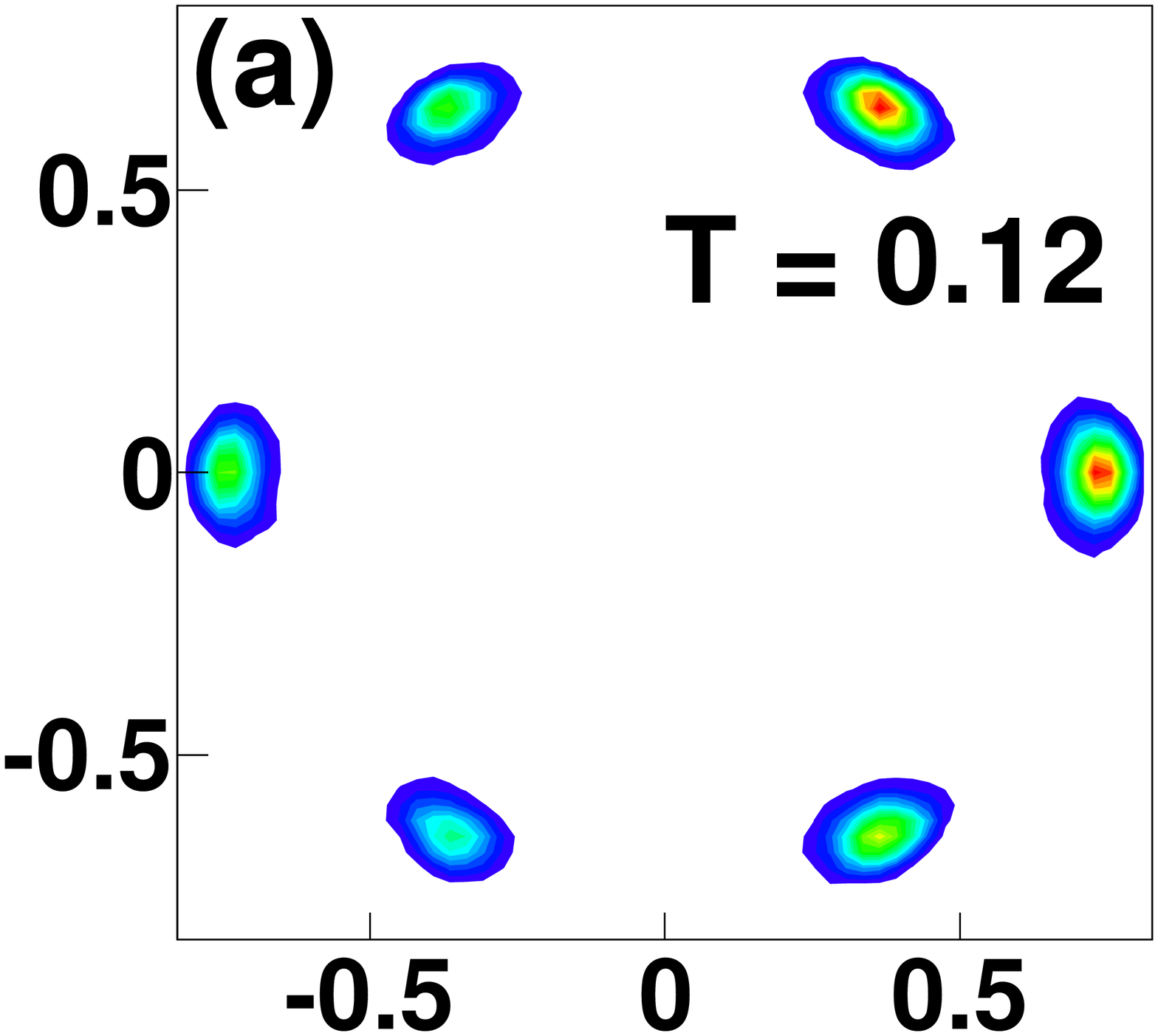}
\includegraphics[width=0.425\columnwidth]{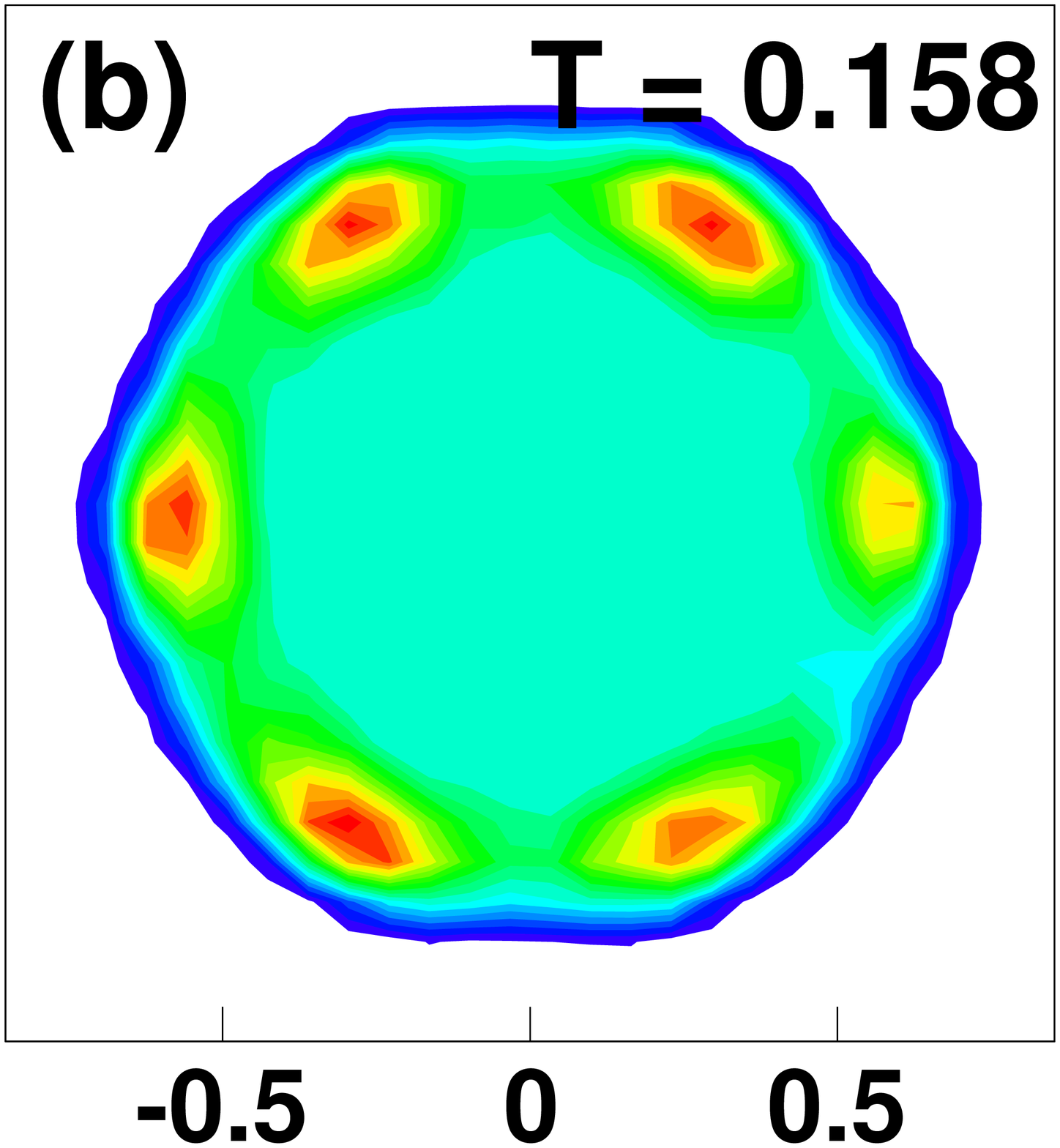}
\includegraphics[width=0.425\columnwidth]{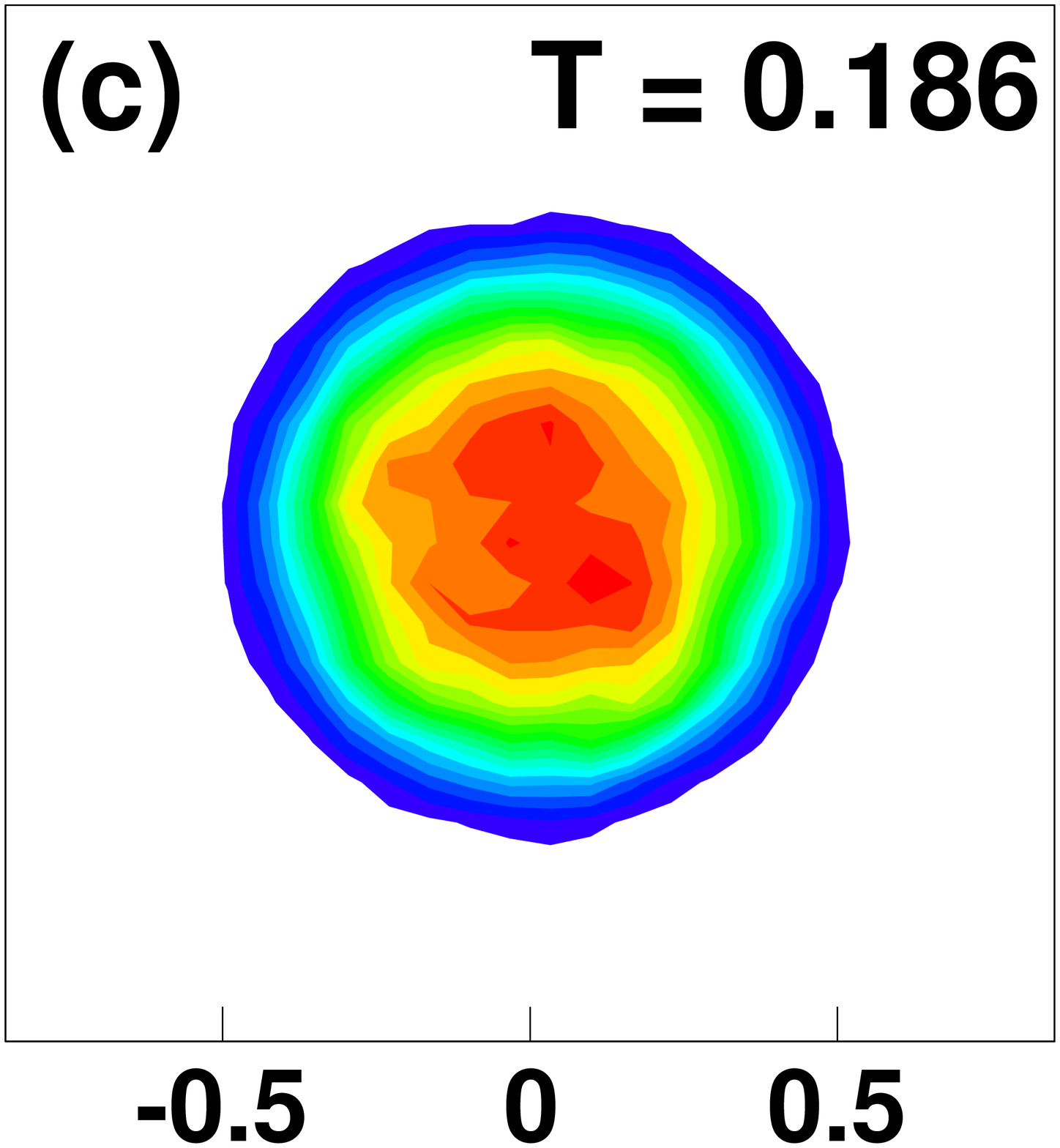}
\includegraphics[width=0.425\columnwidth]{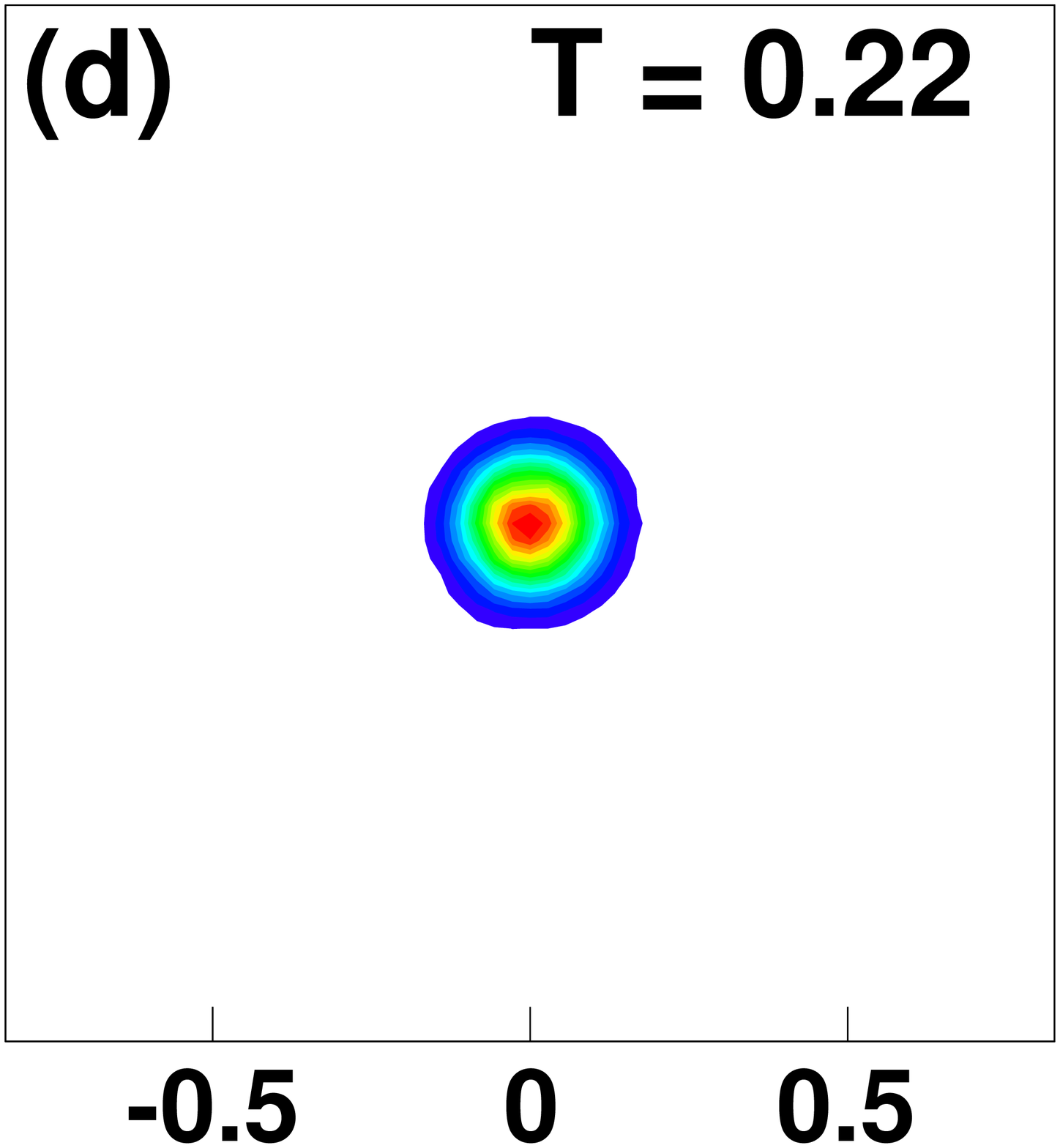}
\includegraphics[width=0.517\columnwidth]{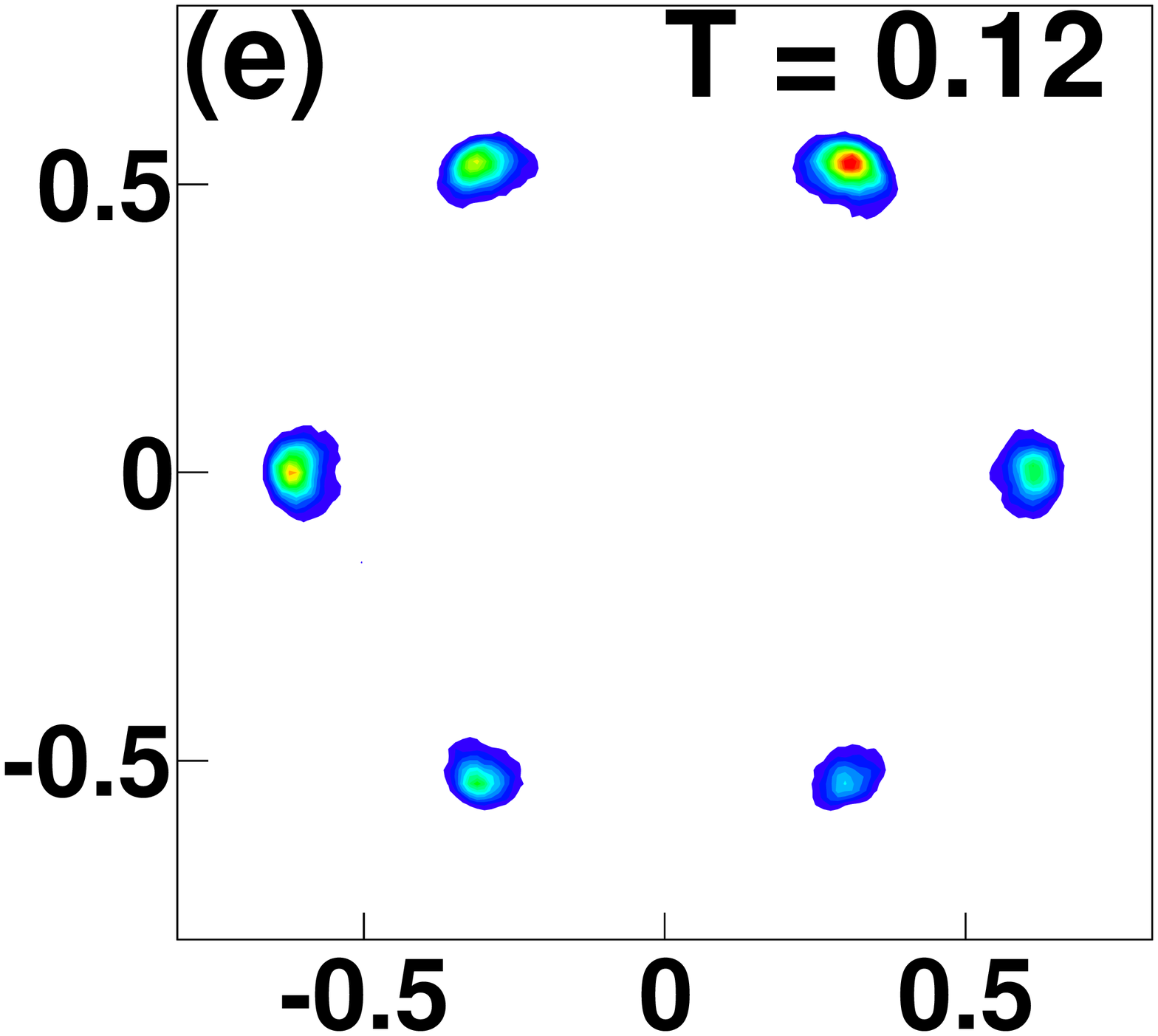}
\includegraphics[width=0.425\columnwidth]{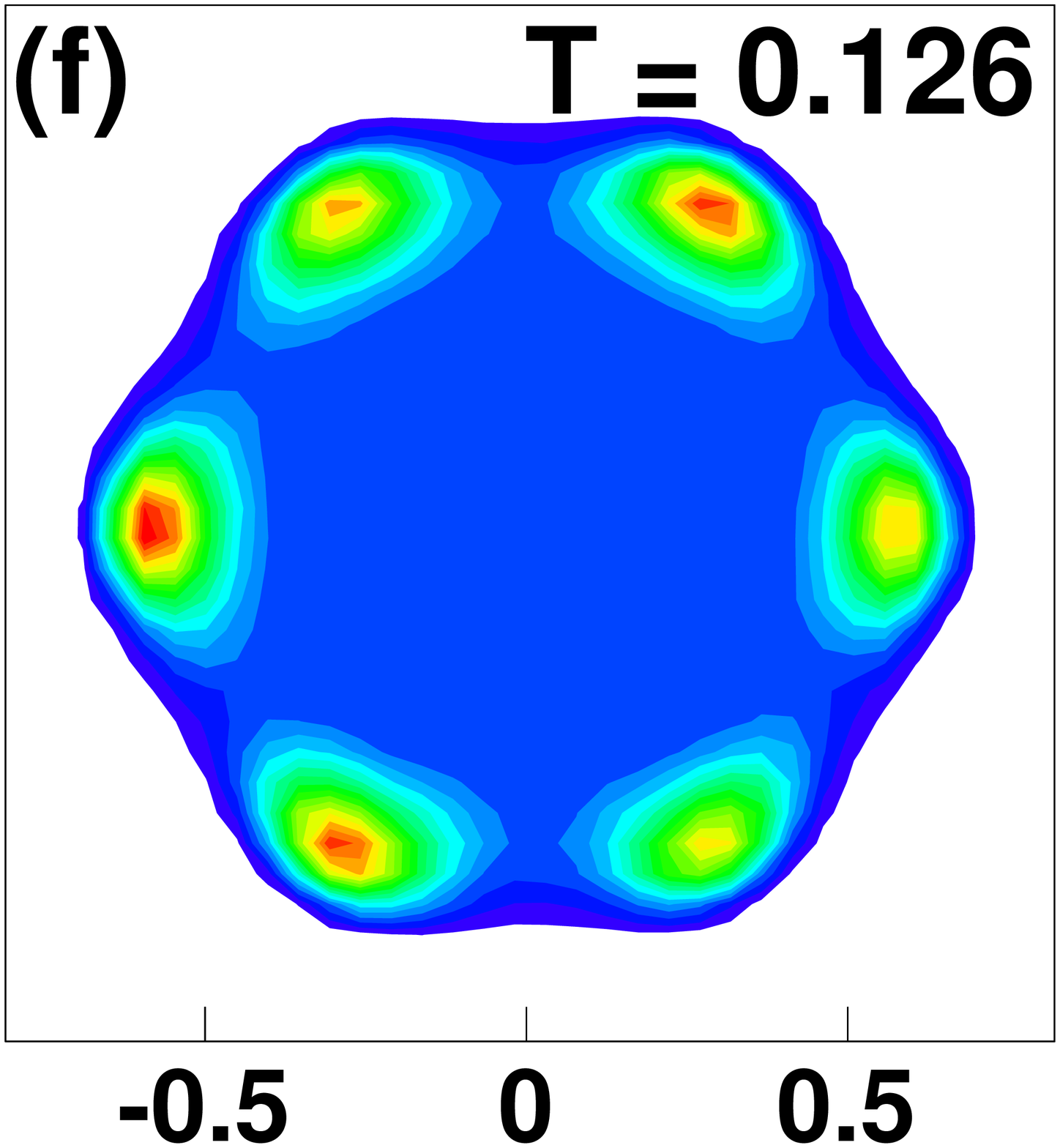}
\includegraphics[width=0.425\columnwidth]{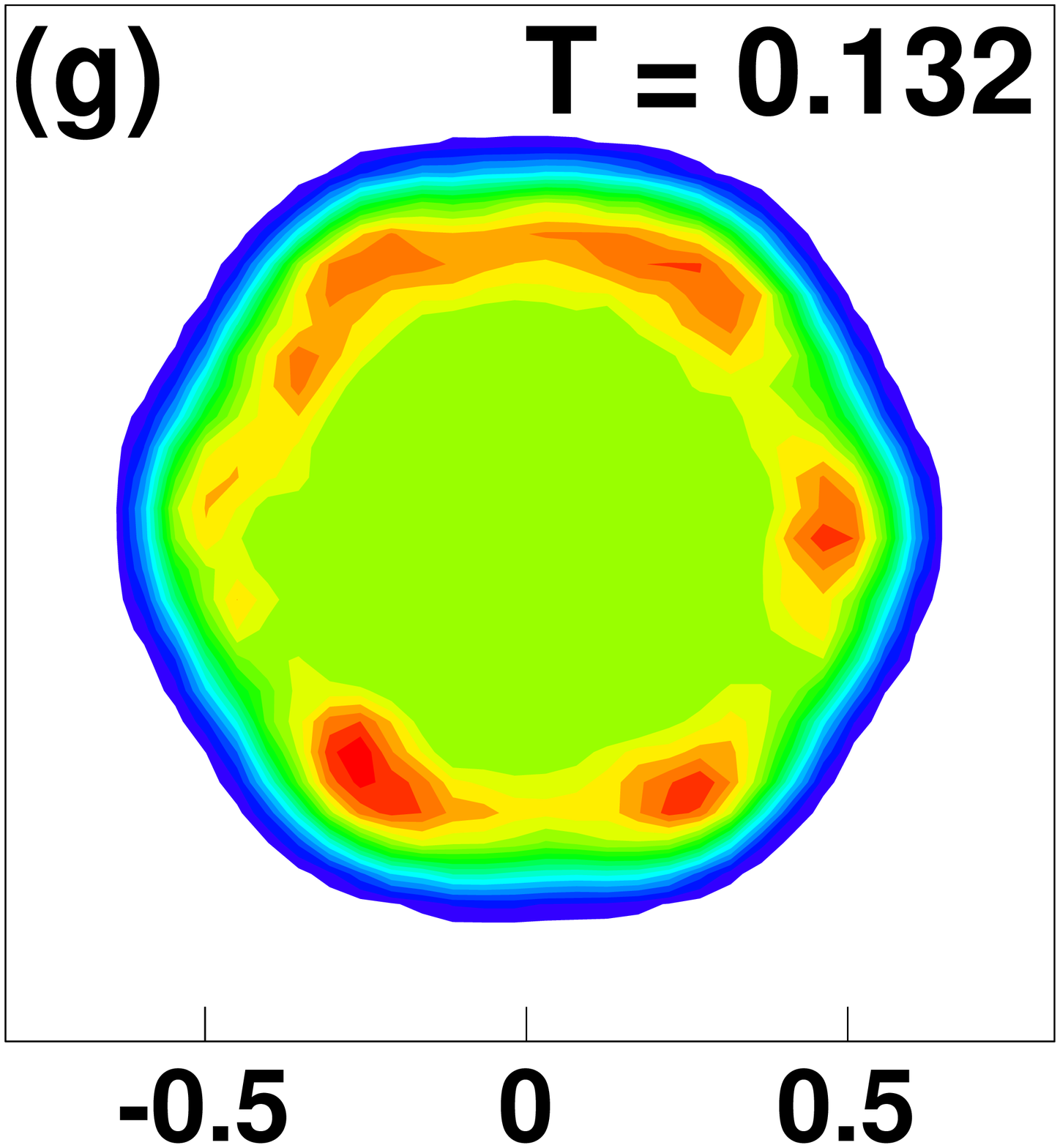}
\includegraphics[width=0.425\columnwidth]{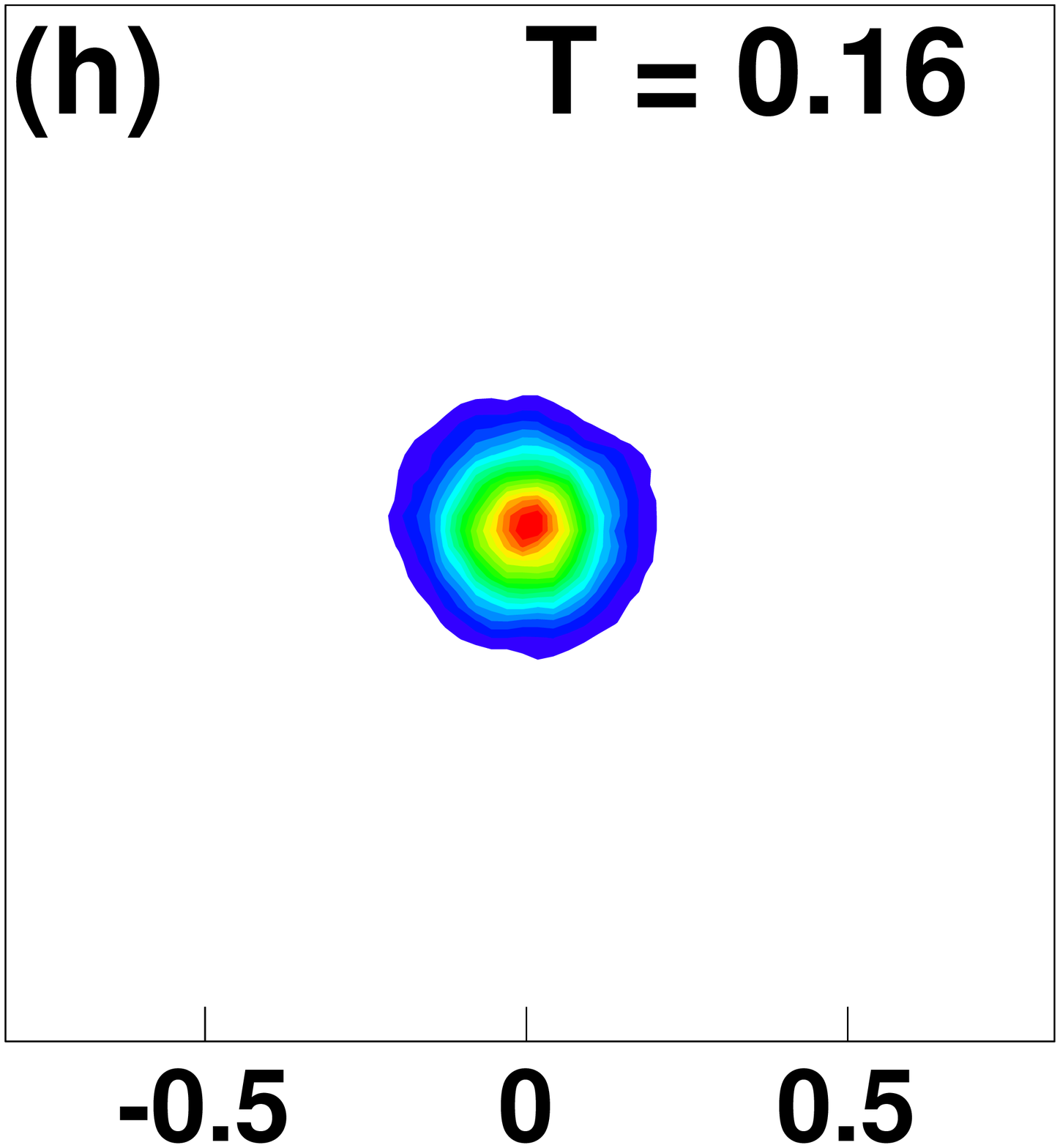}
\caption{
Histograms of the order parameter $m_{N(S)}$, obtained for the system with 2*84*84 spins in the ordered phase, (a) and (e),
in the intermediate phase, (b)-(c) and (f)-(g),
and in the disordered phase, (d) and (h).
Histograms (a)-(d) are  computed for $\alpha=0.25$, and (e)-(h) are  for $\alpha=0.75$.
The histograms are  presented on the complex plane (Re $|m_{N(S)}|$, Im $|m_{N(S)}|$).
%, binned by 175 bins by 175 bins.
The color on the histograms  indicates an arbitrary linear density scale with red being the highest and blue the lowest.
}
\label{fig:histograms}
\end{figure*}
 The six-fold peak structure is observed in the distribution function of the  N\'{e}el and the stripy order parameters for $\alpha=0.25$ (Fig.~\ref{fig:stereo} (b)) and $\alpha=0.75$ (Fig.~\ref{fig:stereo} (d)).
 The peaks correspond to the ordered spin configurations whose  order parameter points along one of the cubic axes.
 In Figs.~\ref{fig:snapshot} (b) and (d) we present snapshots  of  the spin configurations that contribute to the histograms in Figs.~\ref{fig:stereo} (b) and (d).
  In the snapshot of the spin configuration shown in Fig.~\ref{fig:snapshot}(b), the spins that are pointing along the $y-$~direction are antiparallel on the two sublattices.
  This state corresponds to the  N\'{e}el phase with the order parameter directed along $y-$axis.
   In the spin configuration of Fig.~\ref{fig:snapshot} (d),  spins are pointing parallel to the $z-$~direction on two sublattices and  antiparallel on the other two.
  This state corresponds to the  stripy phase with the order parameter directed along $z-$axis.

The projections and snapshots of the order parameter for the two special points $\alpha=0$ and $\alpha=0.5$ are presented in Figs. \ref{fig:stereo} (a) and (c) and Figs. \ref{fig:snapshot} (a) and (c), respectively.
These are the points with continuous symmetry.
As expected, we  see that the projections of the order parameter on the (111) plane (\ref{fig:stereo} (a) and (c)) are more or less equally distributed along a circle.
Note that the multiple "quasi"-peaks appear because  the computations are  performed on a finite size system.
The snapshot for $\alpha=0$ (\ref{fig:snapshot} (a)) shows that a certain direction is chosen in this particular spin configuration but it is not along one of the cubic axes.
This is  not in contradiction with the Mermin Wagner theorem which only precludes the appearance of long range magnetic order at finite temperature.
 Indeed, there  is no long range order for $\alpha=0.0$ because the spins that belong to the same sublattice  are orientated both parallel and antiparallel to the chosen axis.
   A similar situation is observed at $\alpha=0.5$.
 Here, spins on each sublattice point along a certain direction which is not along any cubic axes; thus there is no long range magnetic order at $\alpha=0.5$ as well.

Finally, there is no additional degeneracy of the  order parameter related to its  real-space structure.
Though both  the stripy  and the zigzag order can have  three different bond orderings due to the $120^\circ$ rotational symmetry of the  honeycomb lattice,  the  direction of the order parameter in spin space  is  chosen once the pattern of the bond ordering is chosen.
%Though both  the stripy  and the zigzag order can have  three different bond orderings due to the $120^\circ$ rotational symmetry of the  honeycomb lattice, once the pattern of the bond ordering is chosen, the  direction of the order parameter in spin space  is also chosen.
The three types of bond ordering in real space immediately imply three different types of spin ordering in spin space.
 For example, in the vertical stripy state shown in Fig.~\ref{fig:orders}(c),  the spins are directed along the $z$ axis just like in the snapshot shown in Fig.~\ref{fig:snapshot} (c).
   Due to SOC, the absence of the additional degeneracy can be understood from the  fact that the symmetry transformations act simultaneously on both the spins and the lattice.

\subsection{The critical nature of the intermediate phase. Finite size scaling  analysis.}

Regardless of the specific kind of magnetic order,  the low-T  ordered phase is separated from the high-T paramagnetic phase by  the intermediate phase.
 In our recent study,  we  have shown that  the intermediate phase is a critical phase with two finite-temperature boundaries that correspond to Berezinskii-Kosterlitz-Thouless (BKT) phase transitions.~\cite{price12}

 To describe  the finite temperature properties of the KH model, we use a  projection of the vector order parameter  on the (111) plane.
 The vector  is characterized by both the absolute value and the azimuthal angle of $|m_{N(S)}|$.
  This projection is equivalent to  the $\mathbb{Z}_6$  order parameter, and we write it in a complex form via $m_{N(S)}=\sum_{i=1}^6 |m_{i,N(S)}| e^{ \imath\theta_i}$.
    We chose the phase $\theta$ such that the minimal-energy states of the order parameter, which point along the cubic axes, will be labeled by the values $\theta_i=\pi n_i/3$, $n_i=0,..5$.
\begin{figure}
\includegraphics[width=0.9\columnwidth]{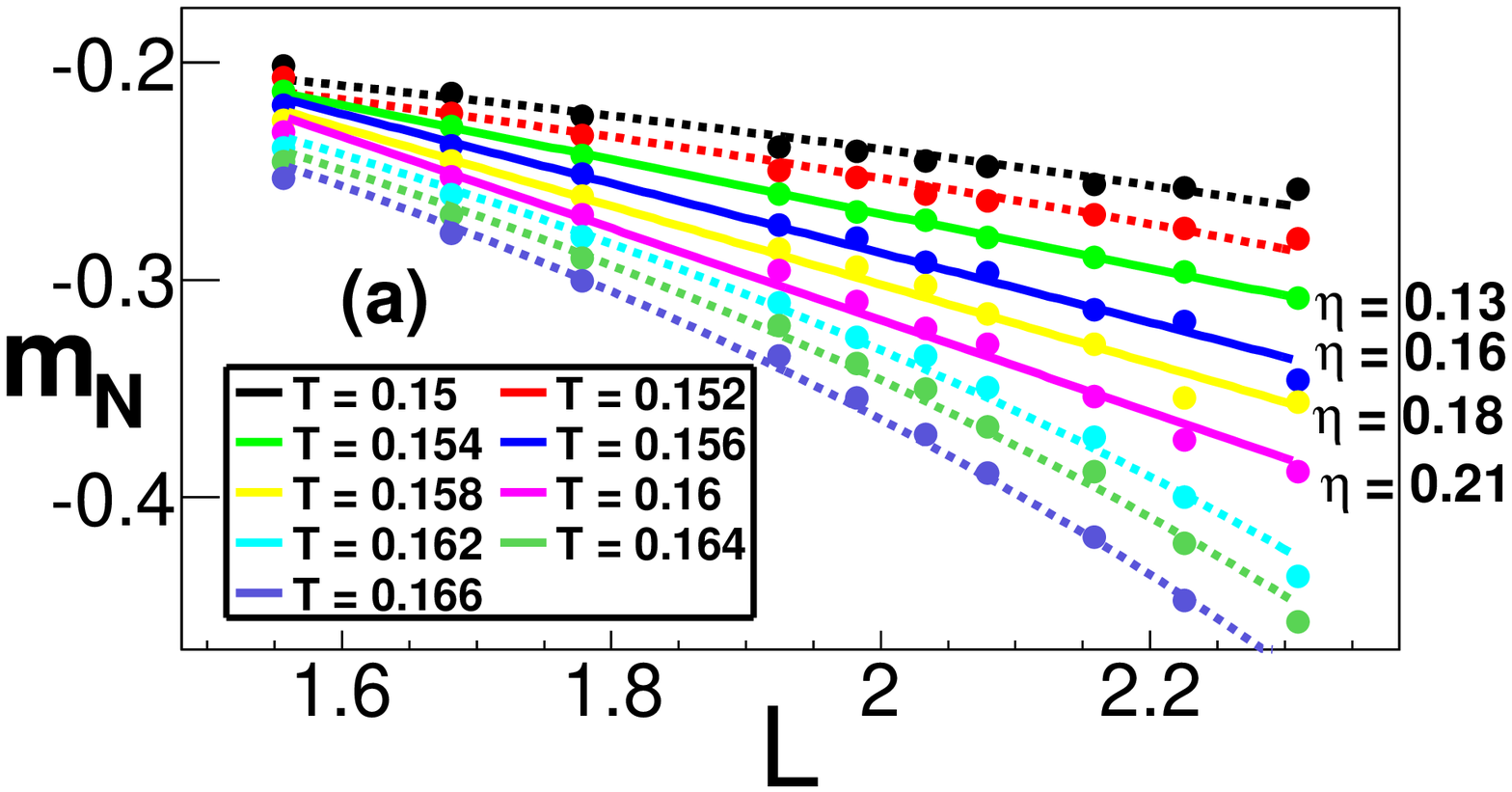}
\includegraphics[width=0.9\columnwidth]{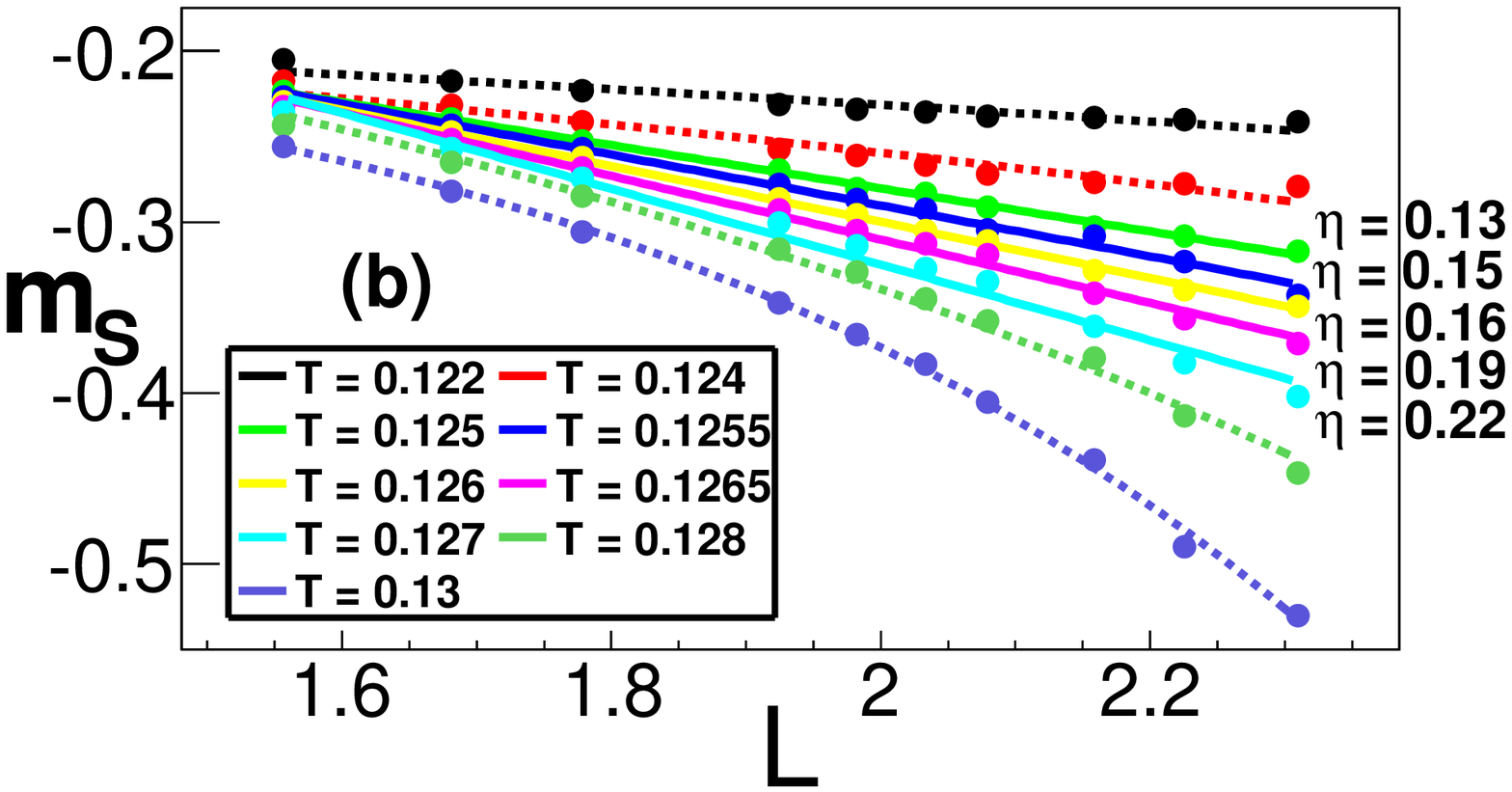}
\caption{
The log-log plots of the order parameter $m_{N(S)}$ as a function of system size $L$ at various temperatures.
The solid curves indicate the linear behavior that corresponds to a power-law dependence, $m_{N(S)}\sim L^{-\eta/2}$,  corresponding to the intermediate critical phase.
The errors  on the linear fit  for the critical exponents $\eta$  are (a)
$\eta(T = 0.154)=0.12 \pm 0.002$, $\eta(T = 0.156)=0.16 \pm 0.006$, $\eta(T = 0.158)=0.18 \pm 0.007$, and
 $\eta(T = 0.160)=0.21 \pm 0.007$ for $\alpha = 0.25$, and
(b) $\eta(T = 0.1250)=0.13 \pm 0.002$, $\eta(T = 0.1255)=0.15 \pm 0.003$, $\eta(T = 0.1260)=0.16 \pm 0.002$, $\eta(T = 0.1265)=0.19 \pm 0.006$ and
 $\eta(T = 0.1270)=0.23 \pm 0.01$ for $\alpha = 0.75$.
 The dashed curves show deviation away from the linear behavior outside the critical phase.
}
\label{fig:loglog}
\end{figure}
Observing the critical phase proved to be challenging for several reasons.
First, within the vicinity of the BKT transition, the critical behavior gives rise  to a very slow, logarithmic convergence to the thermodynamic limit.
We thus had to perform the finite size scaling analysis of our simulations on rather large systems with  $L=84, 96, 108, 120, 144, 168, 204$.
Second,  contrary to the Ising-like spin systems in which  all previous studies of the  six-state clock model's critical phase have been performed, the magnetic degrees of freedom in the classical KH model  are 3D Heisenberg spins which are more strongly affected by thermal fluctuations.
As a result, a large number of sweeps  is needed  to  average out the thermal fluctuations  and  capture the critical behavior.

\begin{figure}
\includegraphics[width=0.45\columnwidth]{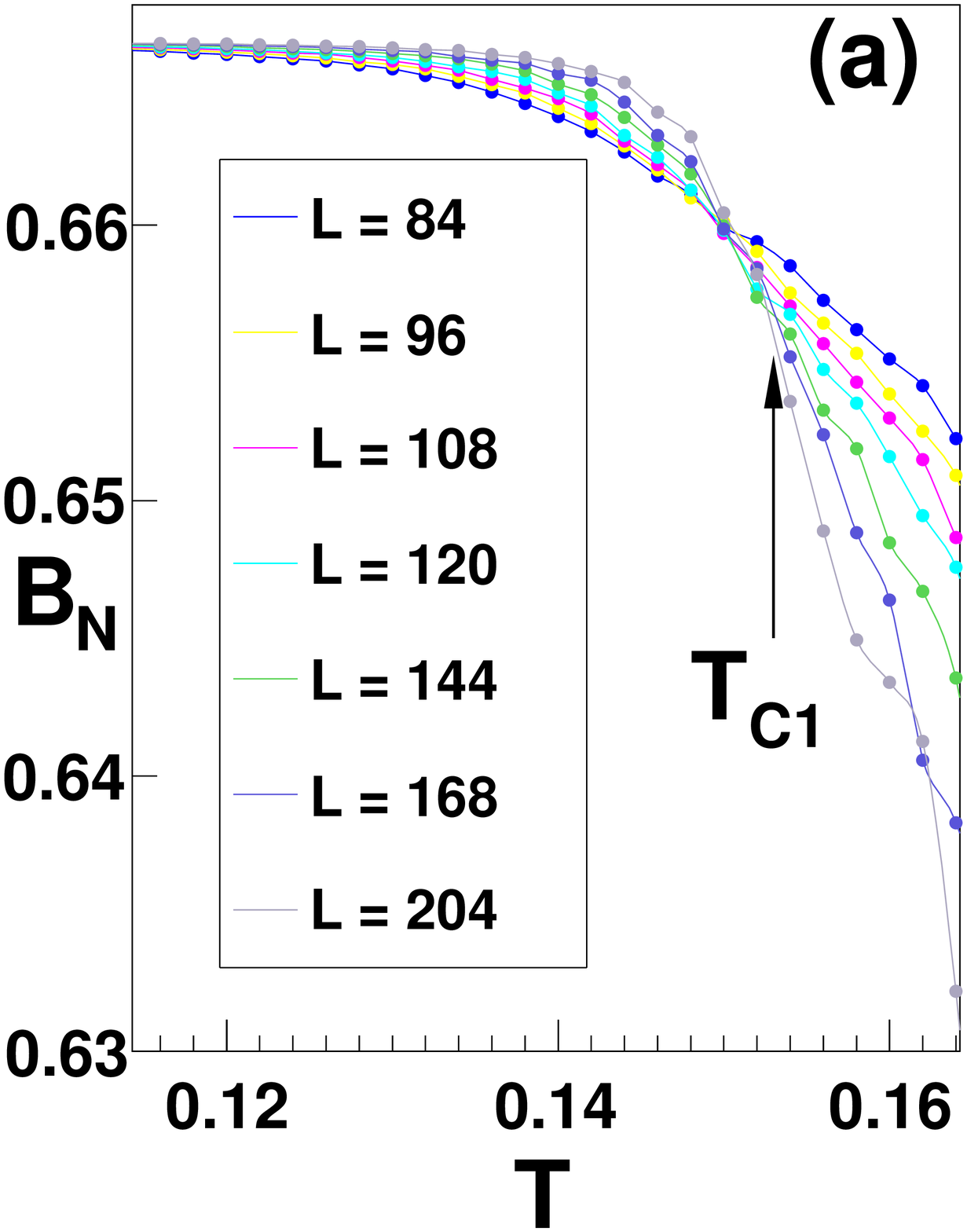}
\includegraphics[width=0.45\columnwidth]{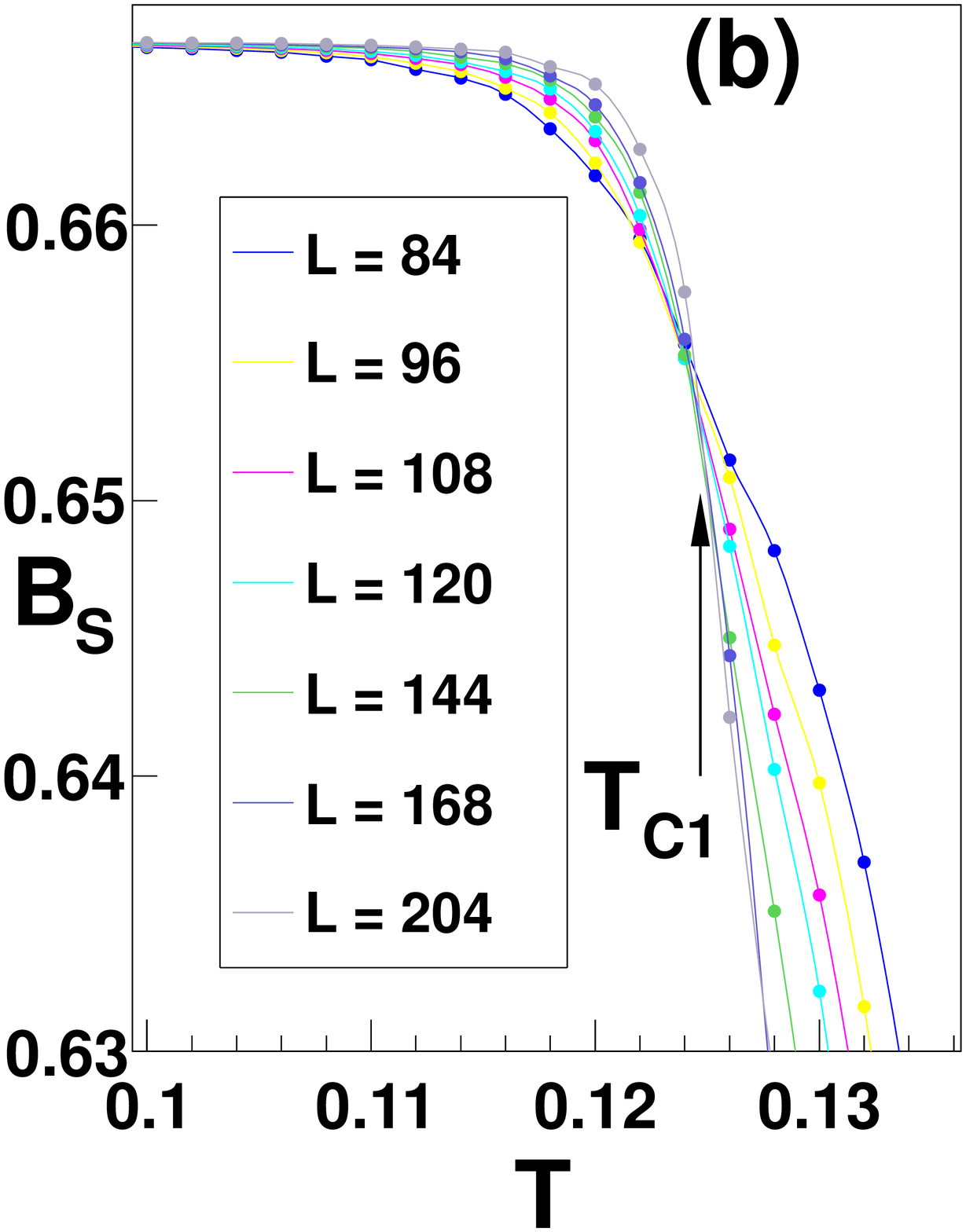}
\caption{
The Binder cumulant as a function of temperature for (a) $\alpha=0.25$ and (b) $\alpha=0.75$.
From  the  crossing points of different Binder's curves, we estimate $T_{c_1}=0.152\pm 0.0005$ and  $T_{c_1}=0.124\pm 0.001$ for  $\alpha=0.25$ and  $\alpha=0.75$, respectively.
}
\label{fig:binder}
\end{figure}

  Evidence of the three-phase structure is clear from the histogram plots  of the complex magnetization, $m_{N(S)}$  (see, Fig.\ref{fig:histograms}).
 As we move from low to high  temperature, we observe a transition from an ordered phase (6 isolated spots) through an intermediate  critical phase (ring distribution) to the disordered phase (uniform distribution around zero).
 The critical phase  has an emergent, continuous $U(1)$ symmetry,\cite{interesting} which is  reminiscent of the intermediate phase of the six-state clock model.\cite{jose77}
   Both inside the critical phase  and at the boundaries, the order parameter exhibits a power law dependence on system size of the form, $m_{N(S)}\sim L^{-\eta/2}$.
 From renormalization group analysis,~\cite{jose77} it is  known that the lower $T_{c_1}$ and the upper $T_{c_2}$ transitions in the  six state clock model  are characterized by the critical exponents $\eta_1=1/9$  and $\eta_2=1/4$, respectively.

\begin{figure*}
\includegraphics[width=0.48\columnwidth]{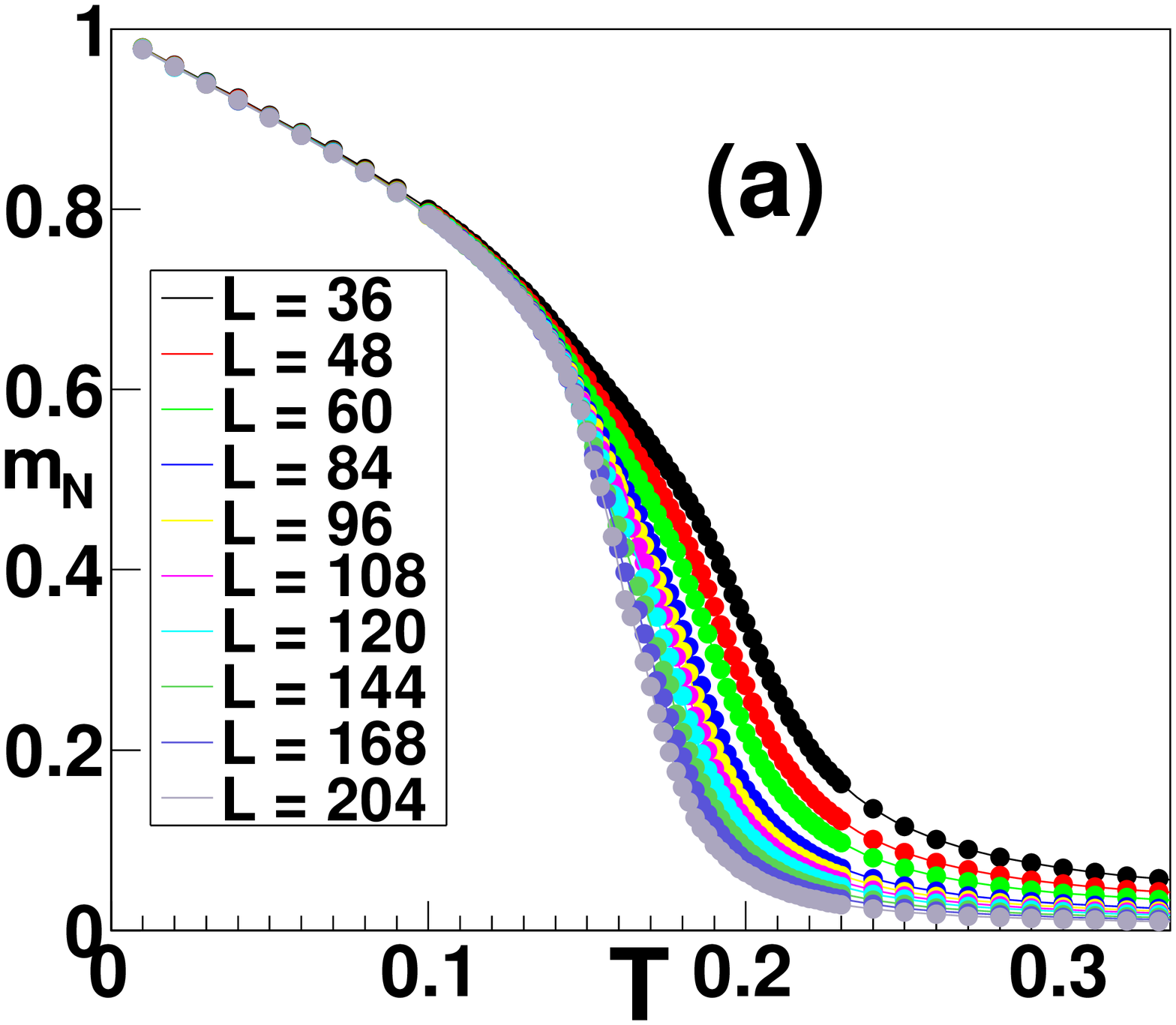}
\includegraphics[width=0.48\columnwidth]{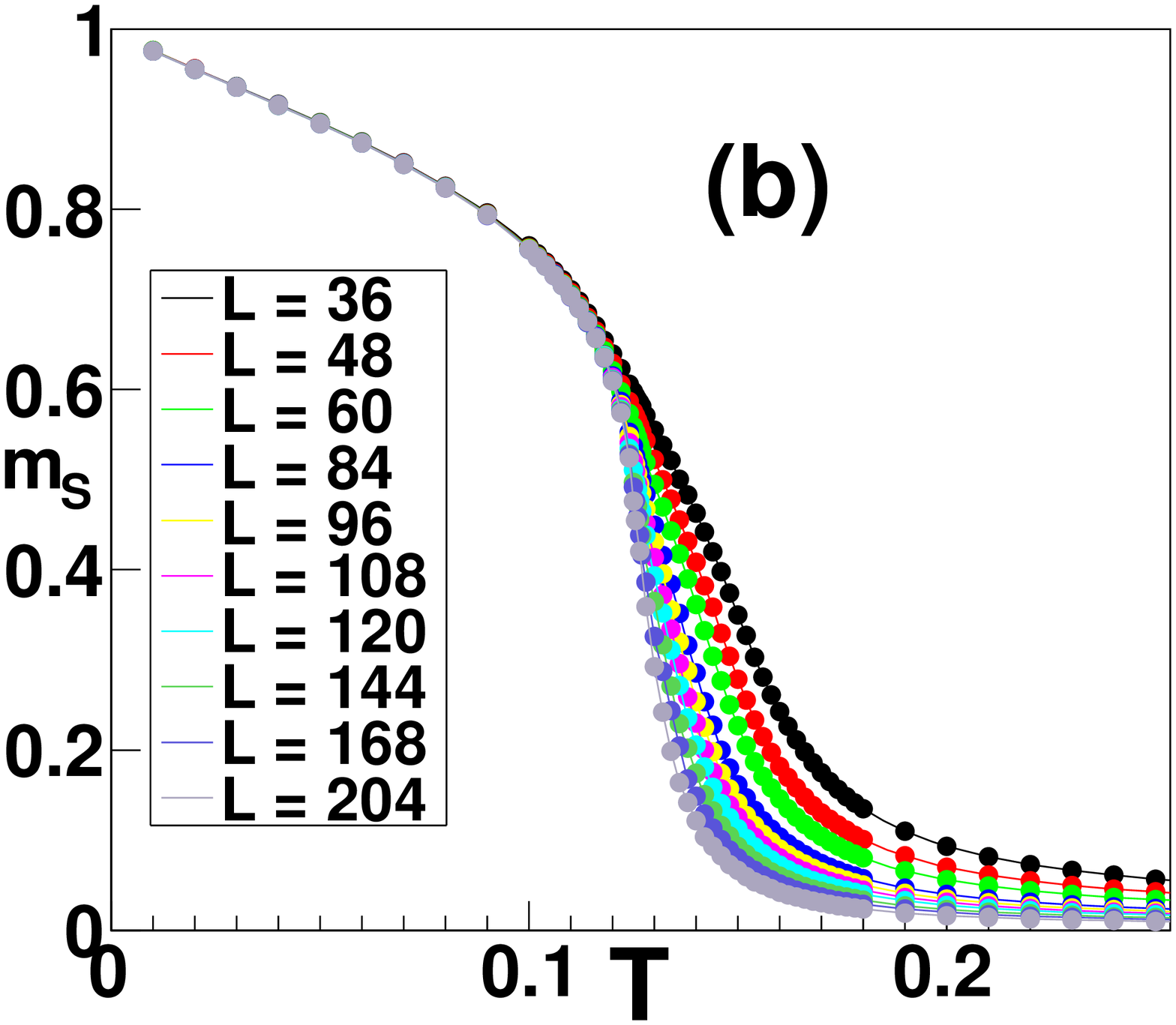}
\includegraphics[width=0.48\columnwidth]{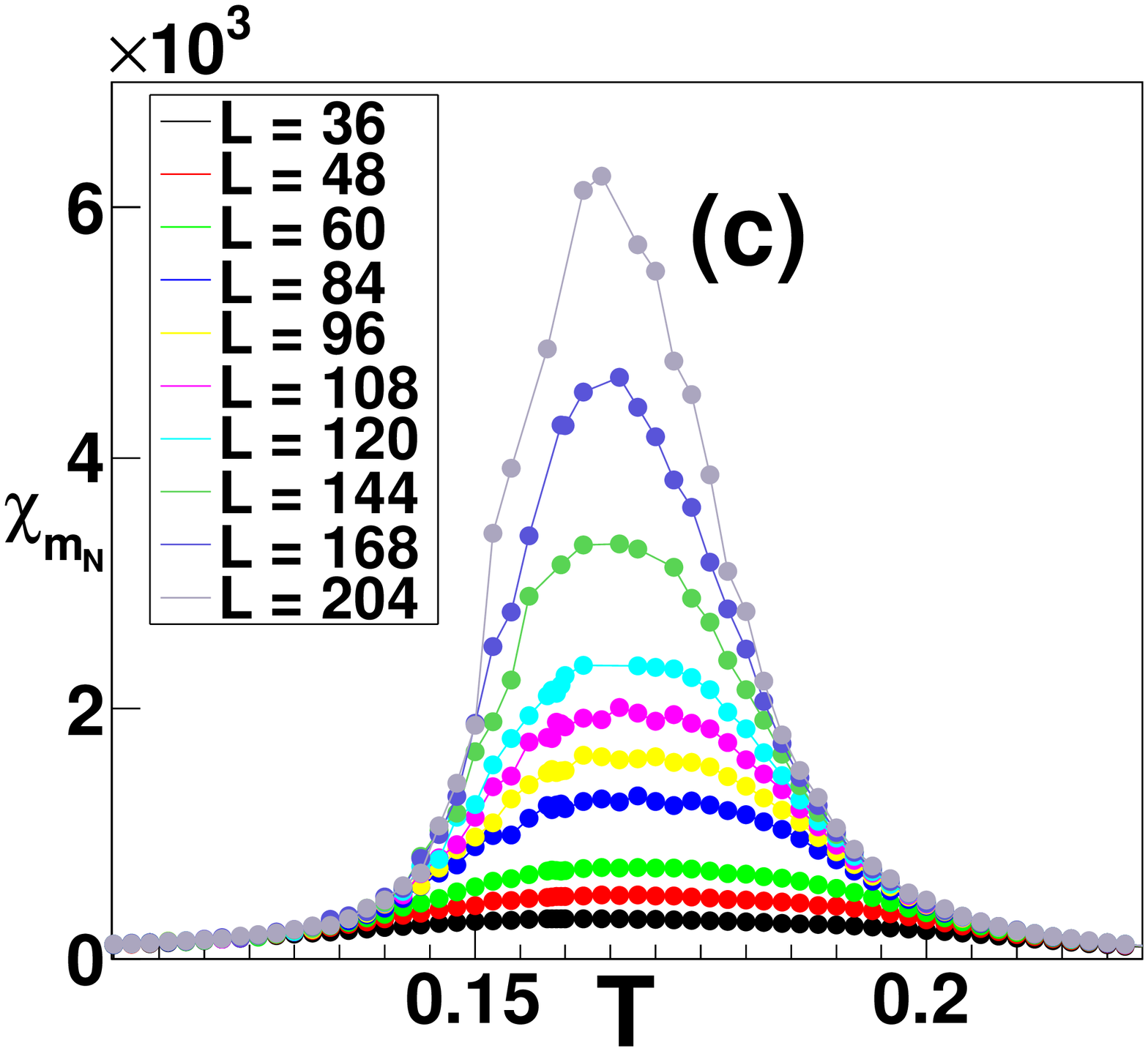}
\includegraphics[width=0.48\columnwidth]{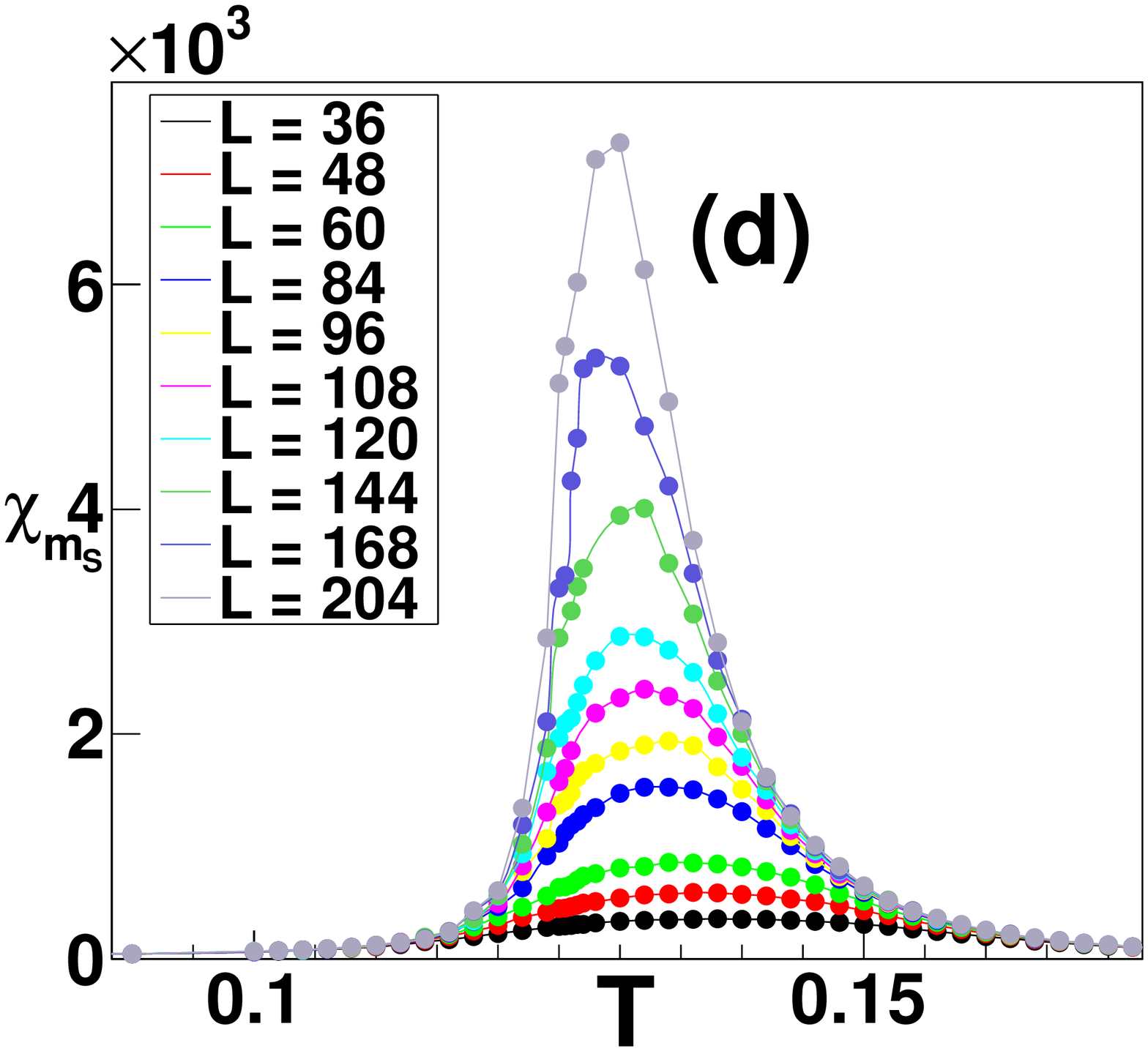}
\includegraphics[width=0.48\columnwidth]{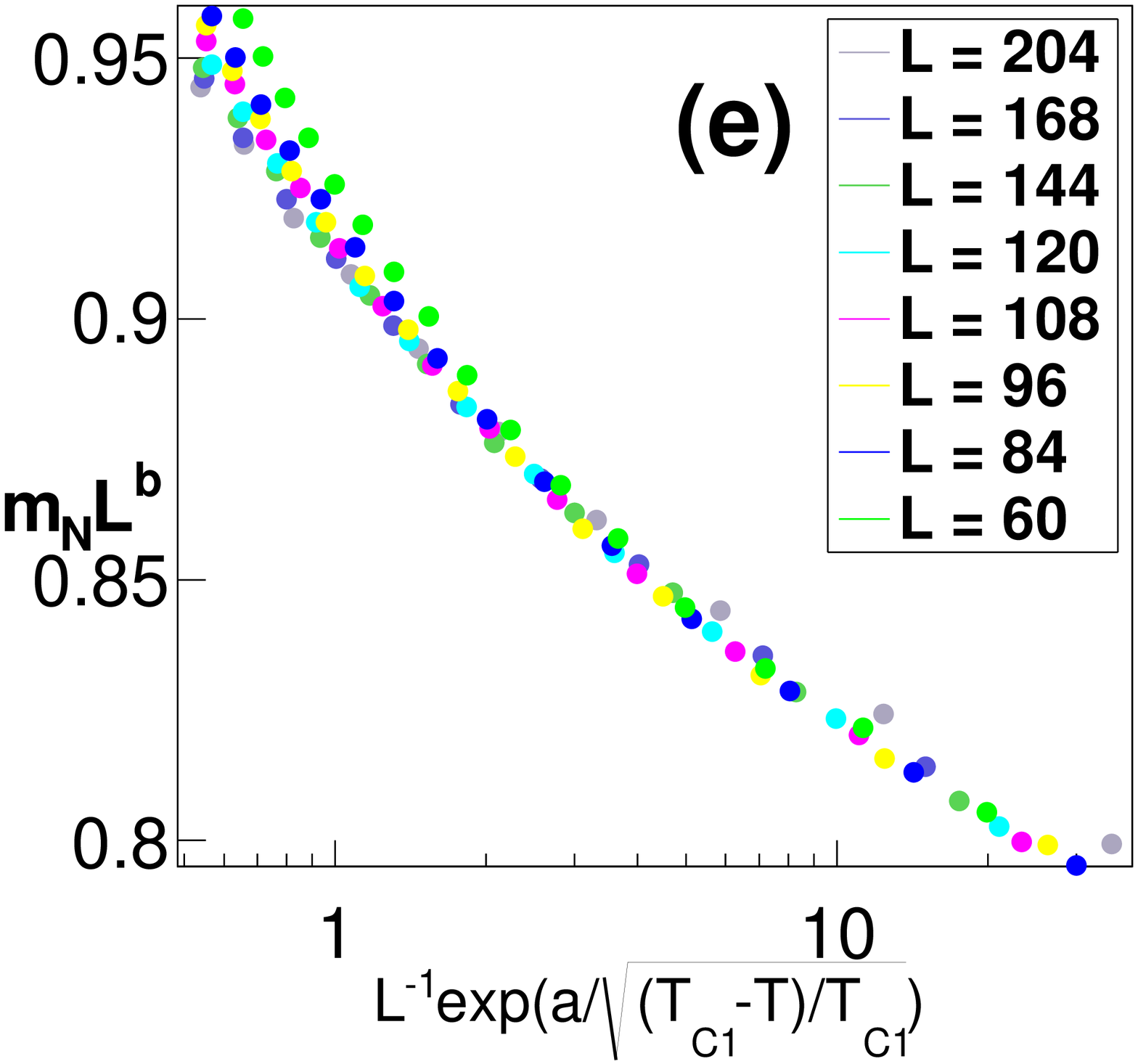}
\includegraphics[width=0.48\columnwidth]{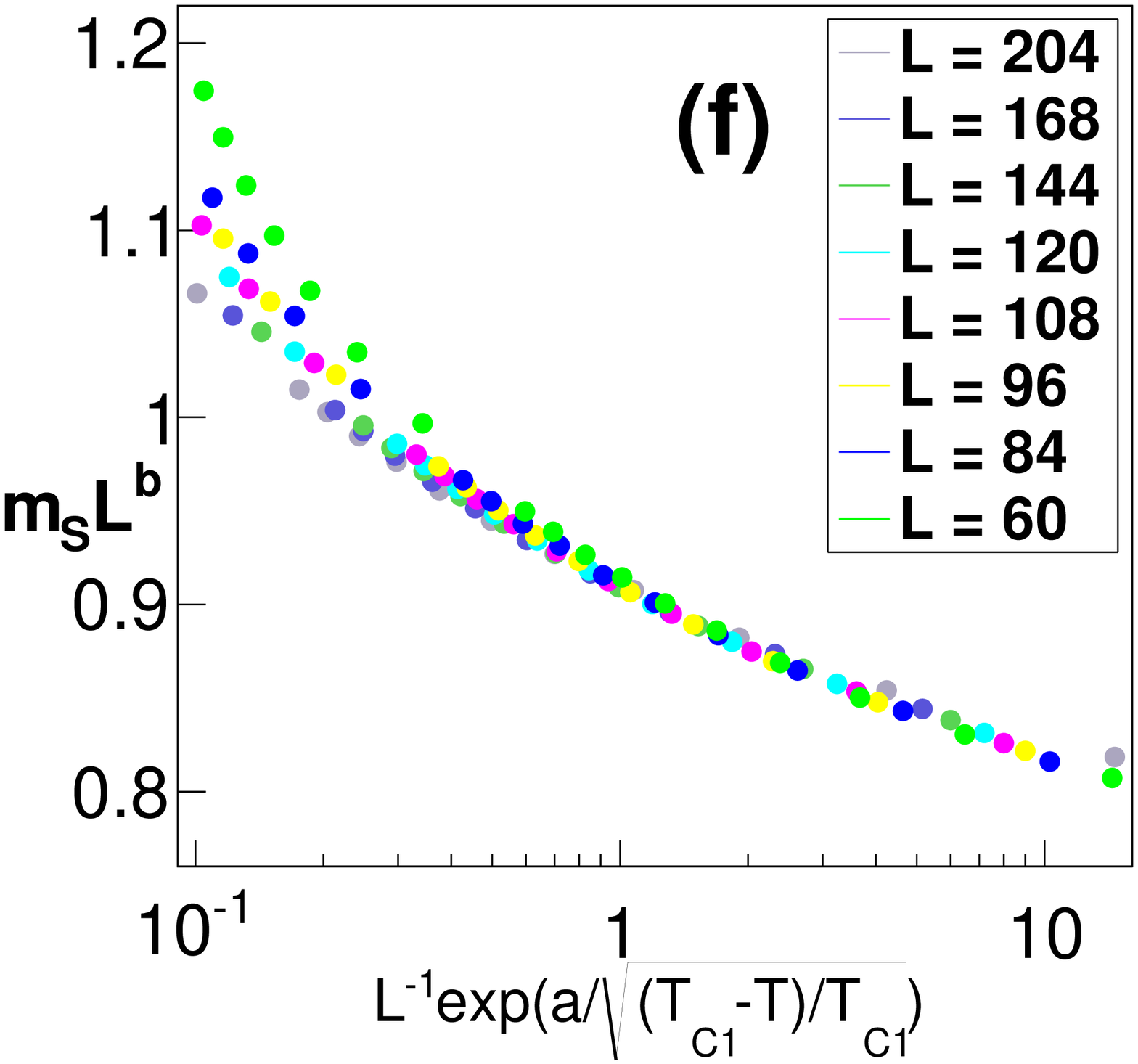}
\includegraphics[width=0.48\columnwidth]{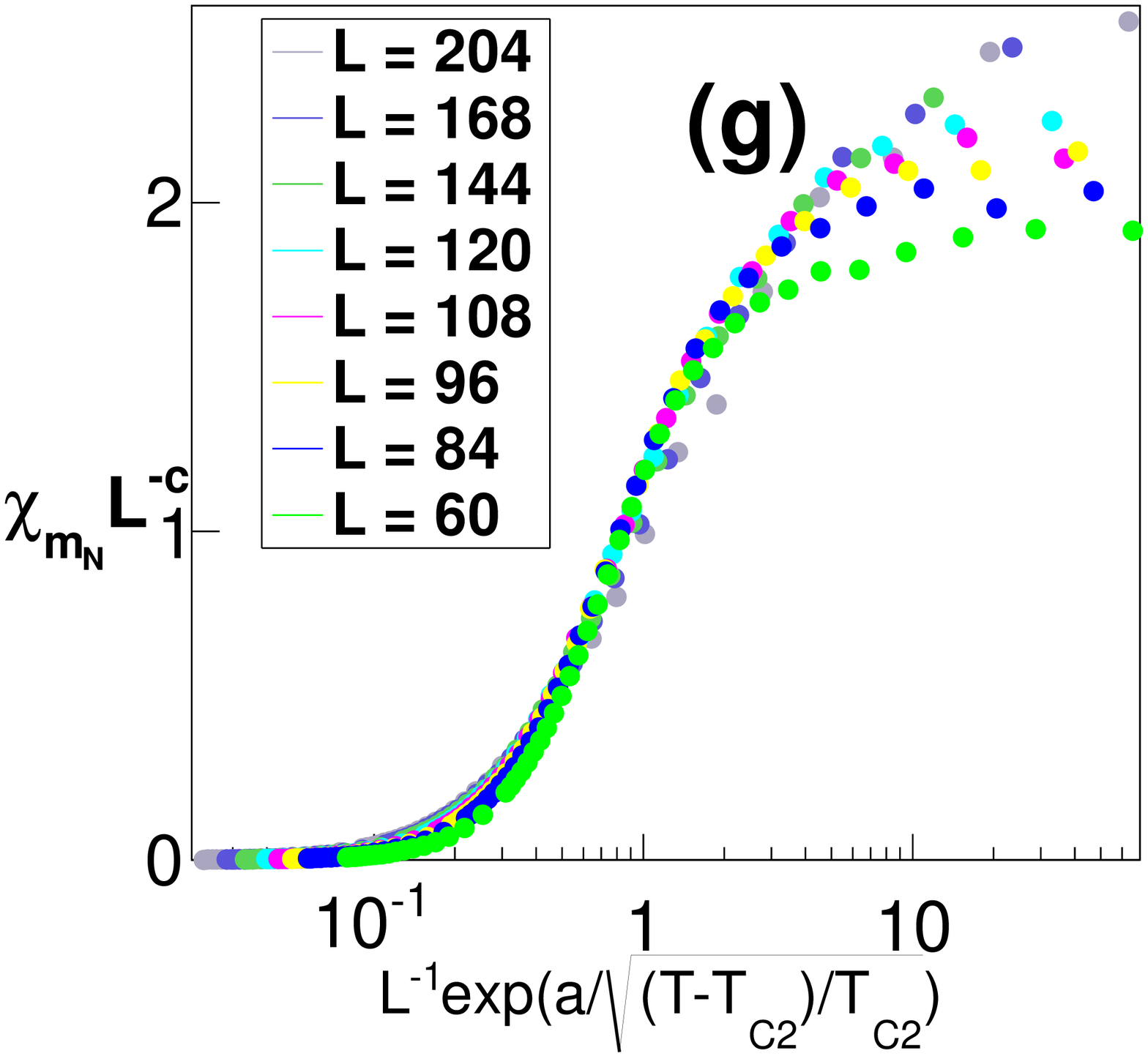}
\includegraphics[width=0.48\columnwidth]{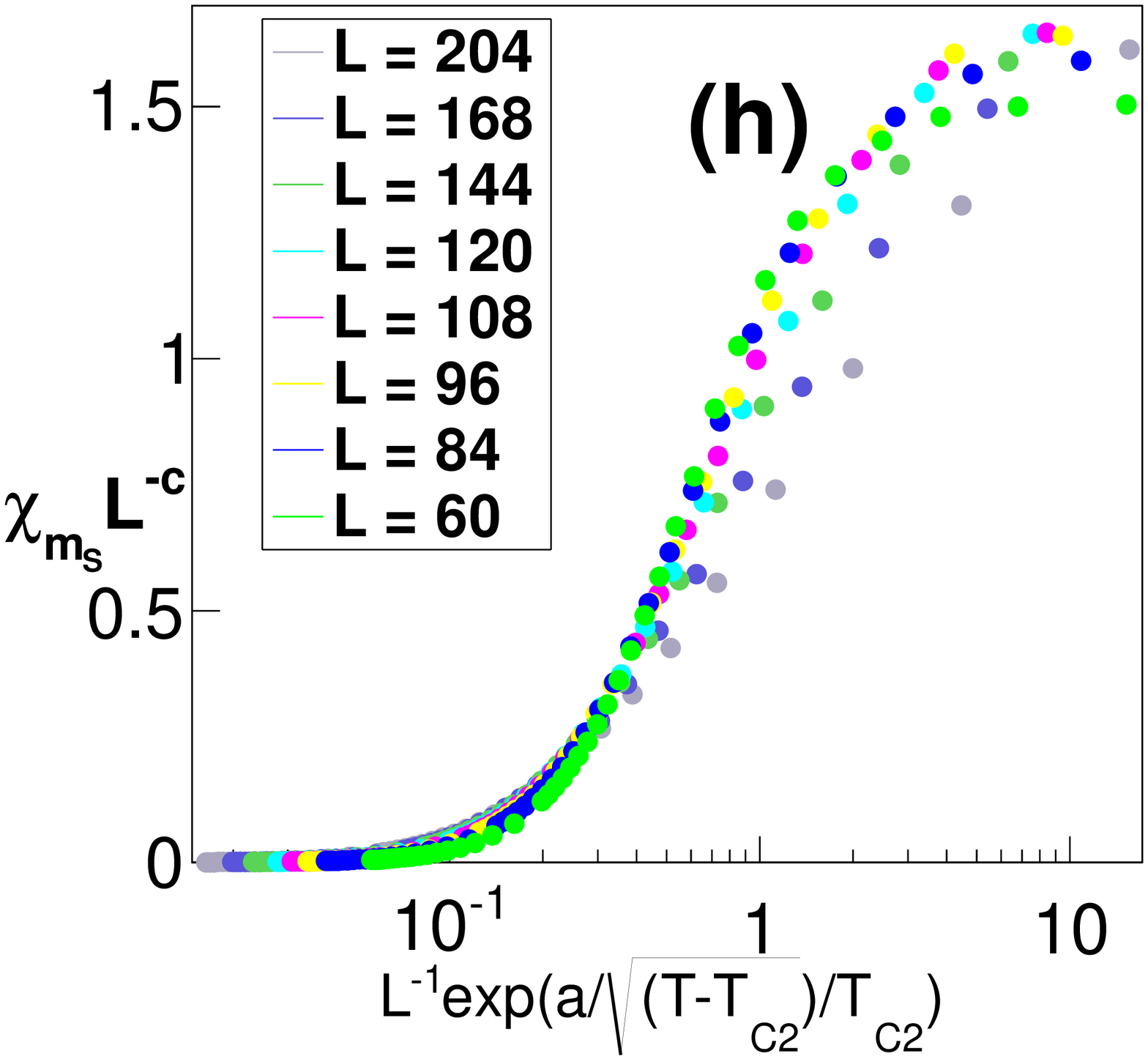}
\caption{Thermodynamic quantities and finite size scaling analysis.
 (a) and (b): The order parameters $m_{N}$ and $m_{S}$ as functions of $T$ for $\alpha=0.25$ and $\alpha=0.75$, respectively.
(c) and (d): The susceptibility, $\chi_{m_{N(S)}}$ (defined in Eq.(\ref{chi})) plotted as a function of $T$ for $\alpha=0.25$ and $\alpha=0.75$, respectively.
(e) and (f): Finite-size scaling of the order parameter
data  in the low-temperature region ($T<T_{c_1}$) for $\alpha=0.25$ and $\alpha=0.75$, respectively.
(h) and (g): Finite-size scaling
of the high-temperature susceptibility data ( $T>T_{c_2}$) for $\alpha=0.25$ and $\alpha=0.75$, respectively.
}
\label{fig:scaling}
\end{figure*}
 Fig.~\ref{fig:loglog}  shows  the log-log plots of the order parameter $m_{N}$  and $m_{S}$  as a function of system size $L$ for different temperatures.
 In accordance  with the power law behavior of the order parameter, the data points   of the log-log  plot show a linear behavior inside the critical phase.
 For $\alpha=0.25$ (Fig.~\ref{fig:loglog} (a)), the log-log  plots of the order parameter as function of $L$  show a linear behavior in the temperature interval between $T_{c_1}\simeq 0.152$ and $T_{c_2}\simeq 0.162$,   with the boundaries characterized by the critical exponents $\eta_1=0.13$ and $\eta_2=0.21$.
 For $\alpha=0.75$, we detected the critical phase in the temperature interval  between $T_{c_1}\simeq 0.125$ ($\eta_1=0.13$) and $T_{c_2}\simeq 0.127$ ($\eta_2=0.22$).
For both $\alpha=0.25$ and $\alpha=0.75$, our estimates for the critical exponents of the  boundaries are in a good agreement with theoretical values  for the six state clock model.

 We  can verify our estimates for the BKT transition temperatures using the Binder cumulants method and further finite-size scaling analysis.~\cite{challa86}
 The Binder cumulants $B_{m_{N(S)}}$, defined in Eq.(\ref{binder}), have a scaling dimension of zero; the crossing  point of the cumulants for different lattice  sizes  provides a reliable estimate for the value of the critical temperature $T_{c_1}$ at which the long range order is destroyed.
 The results for the Binder cumulants $B_{m_{N(S)}}$ are presented in Fig.~\ref{fig:binder}.
The crossing points for $\alpha=0.25$ and $\alpha=0.75$ are at $T_{c_1}=0.153$ and  at $T_{c_1}=0.125$, respectively,  and are in good agreement with estimates obtained from the log-log plots in  Fig.~\ref{fig:loglog}.

 Another way to check the  nature of the critical phase boundaries is by using the data collapse method based on finite size scaling arguments.
 In the BKT transition, the order parameter and its susceptibility exhibit power law behavior, $m_{N(S)}\sim \xi^{-\eta/2}$ and $\chi_{N(S)}\sim \xi^{2-\eta}$, while the correlation length near the critical temperature, $T_c$, diverges as $\xi\sim \exp (a t^{1/2})$, where $a$ is a non-universal constant and $t=|T-T_c|/T_c$ is the reduced temperature.~\cite{kosterlitz74}
  The  finite size scaling analysis is based on the assumption that the singular part of the free energy is a  homogeneous function of system size, $L$, and of correlation length, $\xi$, and only depends on their ratio, $L/\xi$.
Based on this assumption, the finite size scaling behavior of the order parameter and the susceptibility have the following functional forms
 \begin{eqnarray}
 m_{N(S)}= L^{-b} M_{N(S)}\left(\frac{L}{\xi}\right)\\\nonumber
 \chi_{m_{N(S)}}= L^{c} \Xi_{m_{N(S)}}\left(\frac{L}{\xi}\right)~,
 \end{eqnarray}
 where the scaling constants $b=\eta/2$, and  $c=2-\eta$, and $M_{N(S)}$ and $\Xi_{m_{N(S)}}$ are unknown universal functions.
 We plot the variation in temperature of the order parameter   in Fig.\ref{fig:scaling} (a) and (b) and  the susceptibility vs temperature in Fig.\ref{fig:scaling} (c) and (d).
 In Fig.\ref{fig:scaling} (e) and (f) and Fig.\ref{fig:scaling} (h) and (g), we show the scaling plots for  $m_{N(S)} L^b$ as a function of $L^{-1}\exp(a/\sqrt{(T_{c_1}-T)/T_{c_1}})$ and    $\chi_{m_{N(S)}} L^{-c}$ as a function of $L^{-1}\exp(a/\sqrt{(T-T_{c_2})/T_{c_2}})$, respectively.

The finite-size effects in both the order parameter and its susceptibility are striking.
Both the significant finite-size tail of $m_{N(S)}$ that extends  into the intermediate region for $T>T_{c_1}$, and a strong dependence of the position and the height of the susceptibility peak are due to the finite size system.
  These effects are a consequence of an infinite correlation length in the intermediate critical region that looks like  quasi long range order in  finite size systems.
 Nevertheless, the data  points   for different system sizes plotted in their scaled form  collapse  reasonably well onto universal curves that  correspond to the universal functions $M_{N(S)}$ and $\Xi_{m_{N(S)}}$.
 The best data collapse was  obtained for the following scaling parameters:
$a=1.9$, $b=0.056$ and $c=1.45$ and  transition temperatures $T_{c_1}=0.153$ and $T_{c_2}=0.1615$  for $\alpha=0.25$;
$a=1.55$, $b=0.056$ and $c=1.55$ and  transition temperatures $T_{c_1}=0.125$ and $T_{c_2}=0.127$ for $\alpha=0.75$.
  The  values obtained for $b$ and $c$  give the following critical exponents for the lower and the upper boundary of the critical phase: $\eta(T_{c_1})=0.11$, $\eta(T_{c_2})=0.275$, and $\eta(T_{c_1})=0.11$, $\eta(T_{c_2})=0.225$  for $\alpha=0.25$ and $\alpha=0.75$, respectively.

To summarize the discussion  of the numerical data presented above, we can say that we  definitely observe the critical intermediate phase  in the  classical KH model although it slightly deviates from the standard BKT  criticality.
 We believe that the imperfect data collapse is caused  by the presence of three dimensional spin fluctuations of Heisenberg spins and finite sized systems.
 Nevertheless, the smallness of the  discrepancy  and overall  high  quality of the scaling indicate that this effect  is subdominant in the temperature range where the critical behaviour occurs.

\subsection{The finite temperature phase diagram}\label{FTPD}

\begin{figure}
\includegraphics[width=0.99\columnwidth]{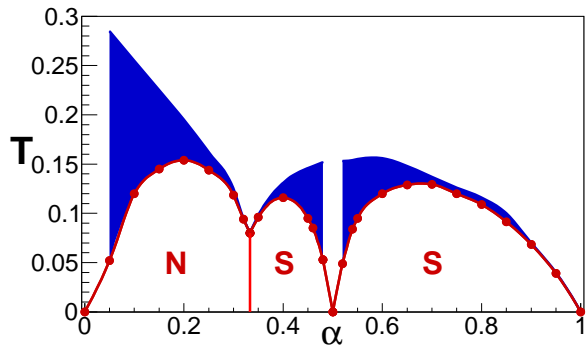}
\caption{Phase diagram of the classical KH model (\ref{ham2}).
 The regions designated by, ``N", and ``S",  are the regions of the phase diagram where the N\'{e}el, and the stripy order persist.
 The vertical line between the N\'{e}el  and the stripy phase at $\alpha=1/3$ denotes the first order phase transition.
Each red circle designates the lower critical temperature, $T_{c_1}$,  for the value of  $\alpha$  for which the model was explicitly simulated; the red line extrapolates between these points.
 For each $\alpha$, the critical value of $T_{c_1}$ was determined through the crossings of Binder's cumulant curves.
The critical  phase is shown in blue.
The upper boundary of the blue critical region was determined by finding the temperature, $T_{c_2}$, for which the value of the critical  exponent $\eta$ exceeds $0.25$.
The blue line extrapolates between each  computed point.
The error bars of the calculated quantities are smaller than the size of the circles representing the data points.
}
\label{fig:phase-diag-alpha}
\end{figure}

In Fig.\ref{fig:phase-diag-alpha} we present the finite temperature phase diagram of the KH model (\ref{ham2}).
We obtained this by tracking the  dependance of the  transition temperatures, $T_{c_1}$ and $T_{c_2}$, on  the strength of the KH model parameter, $\alpha$.
 As discussed in detail in the previous section,  we have used the Binder's cumulant method and finite size  scaling arguments in order to determine critical temperatures.

As expected, only two low-T  magnetically ordered phases are present on the phase diagram - the N\'{e}el phase at $\alpha < 1/3$ and the stripy AFM phase for $1/3<\alpha <1$.
The vertical line that separates  the N\'{e}el and the stripy AFM phases corresponds to a first-order phase transition.
Our numerical results show that  the critical phase is present for all values of $\alpha$ except at the special points $\alpha=0$ and $\alpha=0.5$ where the model has continuous symmetry.
 Thus in accordance with the Mermin-Wagner theorem, the ordered state is destroyed at any non-zero temperature.\cite{mermin66}
As we discussed above for $\alpha=1$, the classical Kitaev model is frustrated. Its classical ground-state structure has a macroscopic degeneracy, and the excitation spectrum of classical  spin excitations has  zero-modes.~\cite{baskaran08,chandra10} Thermal fluctuations of spins are ineffective  to remove this degeneracy, and as a result, the Kitaev model with classical spins shows no order-by-disorder.
 at finite temperatures.

Finally, we note that the way the width of the critical phase disappears  in the vicinity of $\alpha=0$ and $\alpha=0.5$ is different than for  $\alpha=1$.
  In Fig.\ref{fig:phase-diag-alpha}, we can see that the critical phase narrows  very rapidly in the vicinity of  $\alpha=0,0.5$.
 This behavior is related to the fact that  at these particular points the  anisotropy  either does not exist  as at $\alpha=0$ or cancels out  as  at $\alpha=0.5$.
   In the vicinity of these points, the local minima of the free energy that corresponds to different directions of the order parameter  are still separated by finite energy barriers.
   These barriers allow finite temperature ordering through  the order by disorder mechanism.
     At the same time, the width of the critical phase  slowly decreases  when we approach the Kitaev limit and is extremely small  in the vicinity of $\alpha=1$.
    In this limit, the difference  between the classical energies of the stripy, zigzag and ferromagnetic phases  is decreasing with the critical temperature tending to zero.
    At $\alpha=1$, all of the phases are degenerate  (see also Fig.~\ref{fig:cl-energy}) and, as we discussed above, there is no order-by-disorder.

\section{Finite temperature phase diagram  of the extended classical KH model}\label{PDextended}

Here we explore the finite temperature phase diagram of the KH model extended to its full parameter space.
 This  extension takes into account all of the super-exchange processes leading to the coupling between Ir ions and potentially explains the zigzag magnetic order observed in the A$_2$IrO$_3$ compounds.\cite{liu11,singh12}

 There are three physically distinct processes that  determine the ratio between the Kitaev  and the Heisenberg terms in the original KH model.
  Two of the processes involve the virtual hopping of $t_{2g}$ electrons through the  nearest two oxygen ions.
  As it was shown by Jackeli and Khaliulin\cite{jackeli10},  the processes via the upper and lower oxygen ions interfere destructively and the isotropic part of the Hamiltonian exactly vanishes.
  Exchange couplings of neighboring Kramers states on iridium ions   appear due to  the multiplet structure of the excited levels on an Ir ion induced by Hund's and Coulomb couplings.
The third process  involves a direct hopping  between NN $t_{2g}$ orbitals that gives  a finite  Heisenberg term in the KH model.
 All of these processes involve only $t_{2g}$ electrons.
However, there is another possible process:\cite{chaloupka12}  the intersite $t_{2g}\leftrightarrow e_g$ hopping along the 90$^\circ$ Ir-O-Ir paths.
This is the dominant pathway in a 90$^\circ$  geometry since it involves strong $t_{pd\sigma} $ overlap between the $p-$orbitals on oxygen  and the $e_g$ orbitals on iridium ions.
Remarkably, these hopping processes also introduce the Kitaev interaction, but with  a different sign.

 Following Ref.\onlinecite{chaloupka12}, we consider the most general extension of the KH model; both the Kitaev and the Heisenberg interactions can change sign.
  The  Hamiltonian  of the extended model is
\begin{eqnarray}
\label{extended}
    \mathcal{H} =A(2 \sin \phi \sum_{\langle ij \rangle_\gamma} S_i^{\gamma}S_j^{\gamma}+ \cos \phi\sum_{\langle ij \rangle} {\bf S}_i{\bf S}_j),
\end{eqnarray}
where the relative strengths of the effective interaction between Ir magnetic moments are described by  the phase angle, $\phi$, defined in the interval $(0,2\pi)$,  and the overall energy scale, $A =\sqrt{J_K^2+J_H^2}$.
The phase space of the original KH model is covered by the values of $\phi$ from $3\pi/2$ to $2\pi$.

\begin{figure}
\includegraphics[width=0.99\columnwidth]{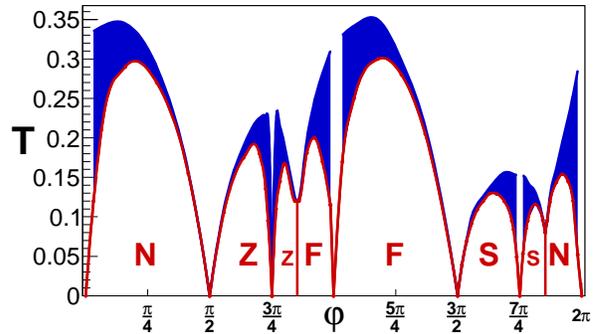}
\caption{
Phase diagram of the extended classical KH model (\ref{extended}).
 The regions designated by, ``N", ``S", ``Z", and ``F" are the regions of the phase diagram where the N\'{e}el, stripy, zigzag, and FM order persist.
Each red circle designates a lower critical temperature $T_{c_1}$  for the value of  $\phi$  for which the model was explicitly simulated.
 For each value of $\phi$, the critical value of $T_{c_1}$ was determined through the crossings of Binder's cumulant curves; the red line extrapolates between these points.
The critical  phase is shown in blue.
The upper boundary of the blue critical region was determined by finding the temperature $T_{c_2}$, for which the value of the critical  exponent $\eta$ exceeds $0.25$.
The blue line extrapolates between each  computed point.
The error bars of the calculated quantities are smaller than the size of the circles representing the data points.
}
\label{fig:phase-diag-phi}
\end{figure}

 Using MC simulations, we investigate the finite temperature properties of this extended model (\ref{extended}) and present the  finite temperature phase diagram in Fig.\ref{fig:phase-diag-phi}.
  In its full parameter space, the extended KH model accommodates  four different  classical phases.
  In addition to the N\'{e}el and the stripy  phases present in the phase diagram (Fig.\ref{fig:phase-diag-alpha}),   the FM and  the zigzag phases appear in Fig.\ref{fig:phase-diag-phi} in a wide range of values of $\phi$.
  The  previously missing  zigzag magnetic order is found  to occupy almost a quarter of the phase space of the extended model.
  Our findings are in agreement with the ground state phase diagram of the quantum model (\ref{extended}) obtained by exact diagonalization.\cite{chaloupka12}
    However, as we already discussed above, the classical model does not support the spin liquid phases present in the quantum model.

  The finite temperature properties of the extended model are very similar to the finite temperature properties of the original KH model.
  Both the transition between the zigzag and the FM order and the transition between the N\'{e}el and the stripy  order are first order phase transitions.
   In the extended parameter space, the  points with a continuous symmetry are $\phi=0,3\pi/4,\pi,7\pi/4$.
  The intermediate phase is present for the whole parameter space of the extended model except for the points where the continuous symmetry is restored and at the special points $\phi=\pi/2$ and $3\pi/2$ because of frustration.
   The  similarities  between the  finite temperature properties of the original and the extended KH model are not surprising.
 The above mentioned  4-sublattice spin transformation,\cite{jackeli10} permits a mapping of the AF stripy phase to the FM phase and of the AF N\'{e}el  phase to the zigzag phase.

  According to the fit of the uniform magnetic susceptibility presented in Ref.~\cite{chaloupka12},  $\phi\simeq 111^\circ \pm 2 ^\circ \simeq 0.62\pi \pm 0.01\pi$ for Na$_2$IrO$_3$  and $\phi\simeq 124^\circ \pm 4 ^\circ\simeq 0.69\pi\pm 0.03\pi$ for Li$_2$IrO$_3$.
   Using these values, we get the following estimates for the N\'{e}el temperatures: $T_{c_1}\simeq 0.16 $  which is equivalent to about 17.7 K for  Na$_2$IrO$_3$,  and $T_{c_1}\simeq 0.19 $  which is equivalent to about 21 K for  Li$_2$IrO$_3$.
  Both  values are close to  the experimental value $T_N\simeq 15$ K obtained for both Na$_2$IrO$_3$ and Li$_2$IrO$_3$ compounds.\cite{singh10,singh12,ye12}
%0.152*4.17*4*11.6/1.75=16.8 K
%(3-5alpha)/3*S^2=(3-1.25)/3=1.75/4 should be 4.17. Thus our energy unit is 12.7 meV
 Our estimates for the upper boundary of the critical phase is $T_{c_2}=0.18$ for Na$_2$IrO$_3$ and  $T_{c_2}=0.22$  for Li$_2$IrO$_3$  which  is equivalent to about 20 K and  24.5 K, respectively.
 This gives the estimates for the width of the critical phase to be about 2.3 K and 3.5 K for Na$_2$IrO$_3$ and Li$_2$IrO$_3$, respectively.
 We  also note that our  prediction for two phase transitions are in good  agreement with  the specific heat data  obtained for Na$_2$IrO$_3$  and Li$_2$IrO$_3$.\cite{singh12, ye12}
   In both works the specific heat data  show the anomaly above the N\'{e}el  temperature to be at about $T=18-20$K, indicating the presence of a higher  temperature transition.
   In addition, just above  $T_N$ the entropy obtained by integrating the $\Delta C/T$ versus $T$ data for both  Na$_2$IrO$_3$  and Li$_2$IrO$_3$ is significantly reduced.~\cite{singh10,singh12}.
 Both observations are in line with our prediction of the critical phase above the magnetically ordered phase.

\section{Classical honeycomb Heisenberg antiferromagnet   with  a cubic anisotropy}\label{cubic}

 To complete this study,  we also investigate the nature of the finite-temperature phase transitions of the  honeycomb Heisenberg antiferromagnet in which the cubic anisotropy  is included explicitly.
In this case, the model is defined   by the Hamiltonian
\begin{eqnarray}
\label{hamcubic}
    \mathcal{H}_c =J_H\sum_{\langle ij \rangle} {\bf S}_i{\bf S}_j-D \sum_{i} \left((S_i^{x})^4+(S_i^{y})^4+(S^{z})^4\right)~,
\end{eqnarray}
where $D$ denotes the strength of the cubic anisotropy.
The  positive sign of the anisotropy, $D>0$,  determines  that spins tend to align along the cubic axes and not  in the diagonal directions of the lattice which  corresponds to $D<0$.
Because of the cubic anisotropy, the model (\ref{hamcubic}) shows no full rotational  symmetry.
The spin projections in the  (111) plane again will  favor six directions as in the case of the KH model.
Here we explore by MC simulations whether or not the finite temperature properties of models (\ref{ham1}) and (\ref{hamcubic}) are similar.

Let us discuss the   thermodynamic properties of the  model (\ref{hamcubic}).
At $T=0$ and in the absence of the anisotropy, the ground state of the model (\ref{hamcubic})  is  determined by the  N\'{e}el  order parameter $\mathbf{N}$.
 In the presence of the anisotropy, $D$,  the ordered state will survive  until some critical temperature.
 At high temperature, the thermal fluctuations of spins will destroy the magnetic  order and the system will be in the disordered paramagnetic state.
  As in the KH model, in order to reach the ordered state from  high temperature, the system has to break the discrete $\mathbb{Z}_6$ symmetry of the Heisenberg honeycomb model with cubic anisotropy.
  In the case of the KH model, the $\mathbb{Z}_6$ symmetry was broken through the critical intermediate phase.
   In our simulations of (\ref{hamcubic}), we did not observe any critical phase; instead,  the $\mathbb{Z}_6$ symmetry is broken in  the following two steps.
 After lowering the  temperature, the $\mathbb{Z}_3$ symmetry is broken first at $T=T_{c_2}$; the system still remains paramagnetic since the  expectation value of $\langle \mathbf {N}\rangle =0$.
  In the second step, the remaining $\mathbb{Z}_2$ symmetry is broken at the temperature $T=T_{c_1}<T_{c_2}$ and the system acquires long-range magnetic order.

\begin{figure*}
\includegraphics[width=0.48\columnwidth]{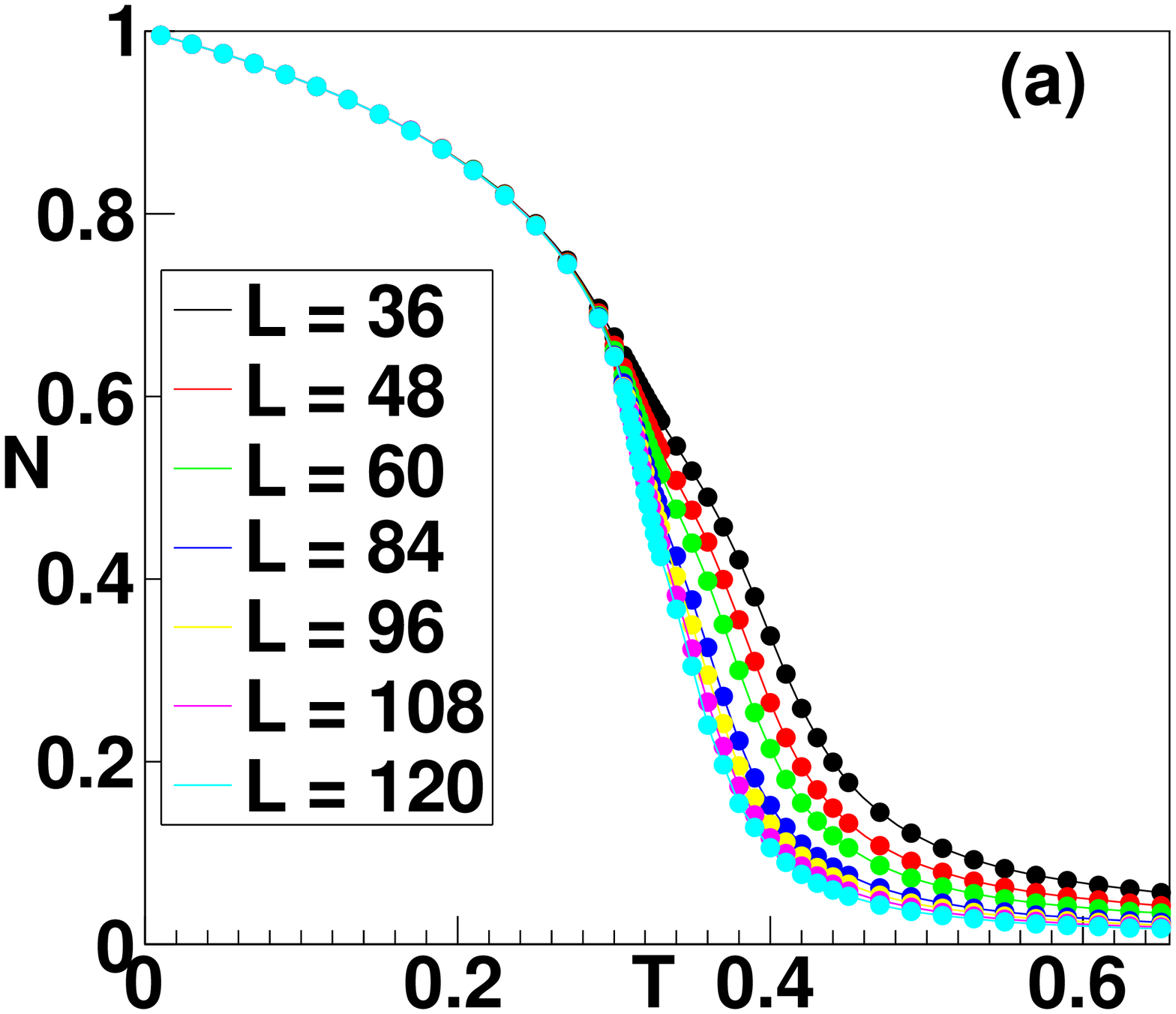}
\includegraphics[width=0.48\columnwidth]{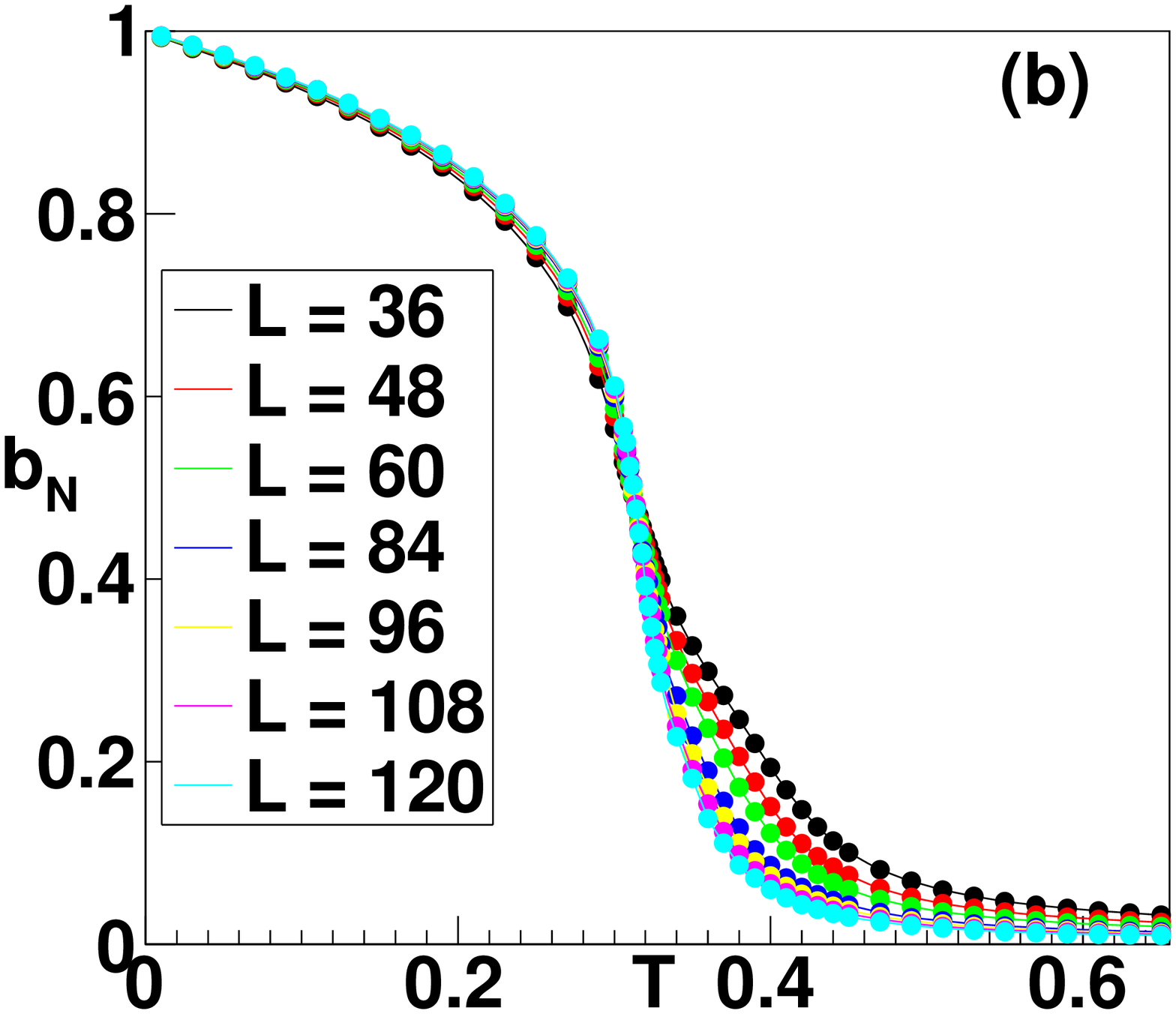}
\includegraphics[width=0.48\columnwidth]{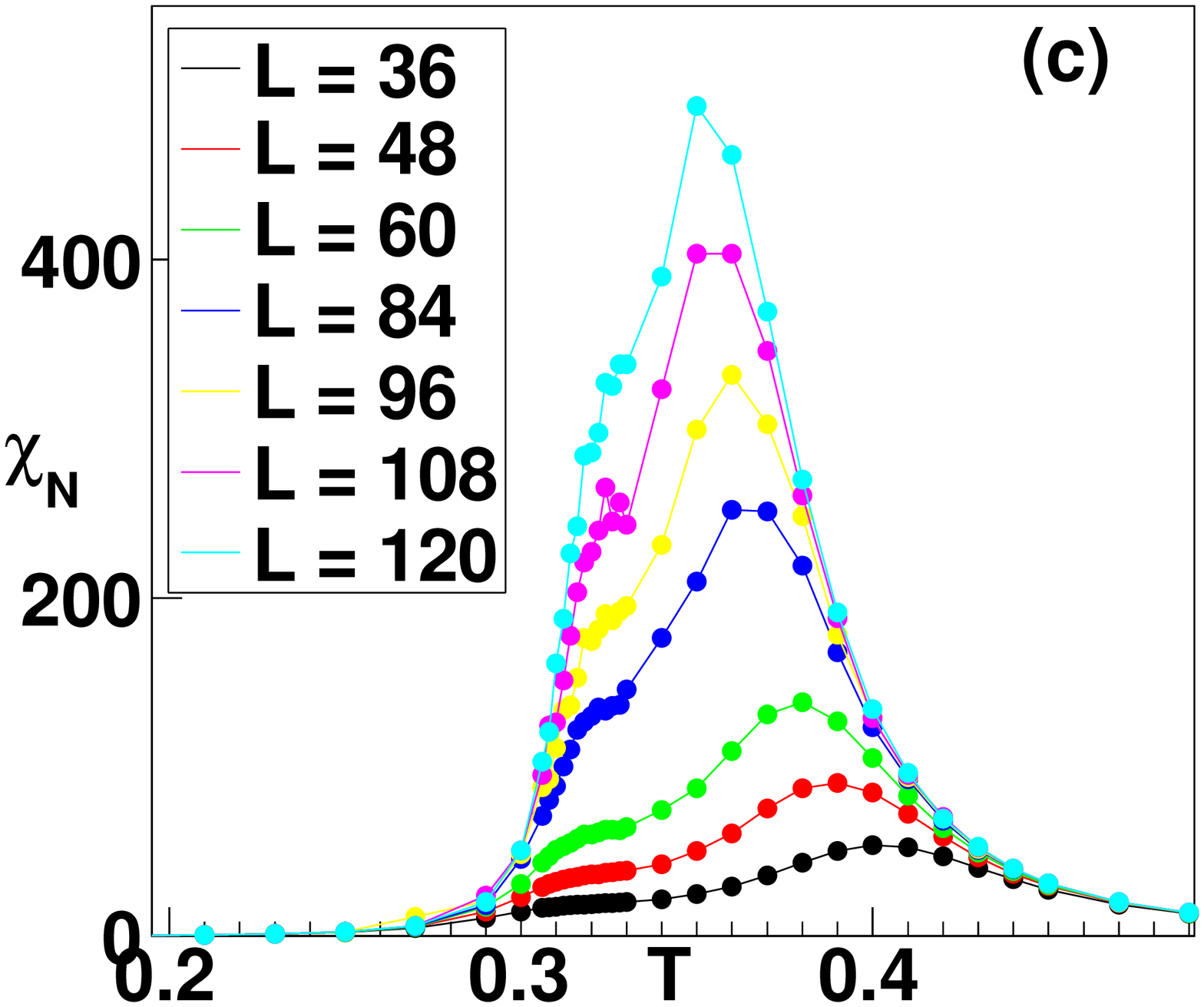}
\includegraphics[width=0.48\columnwidth]{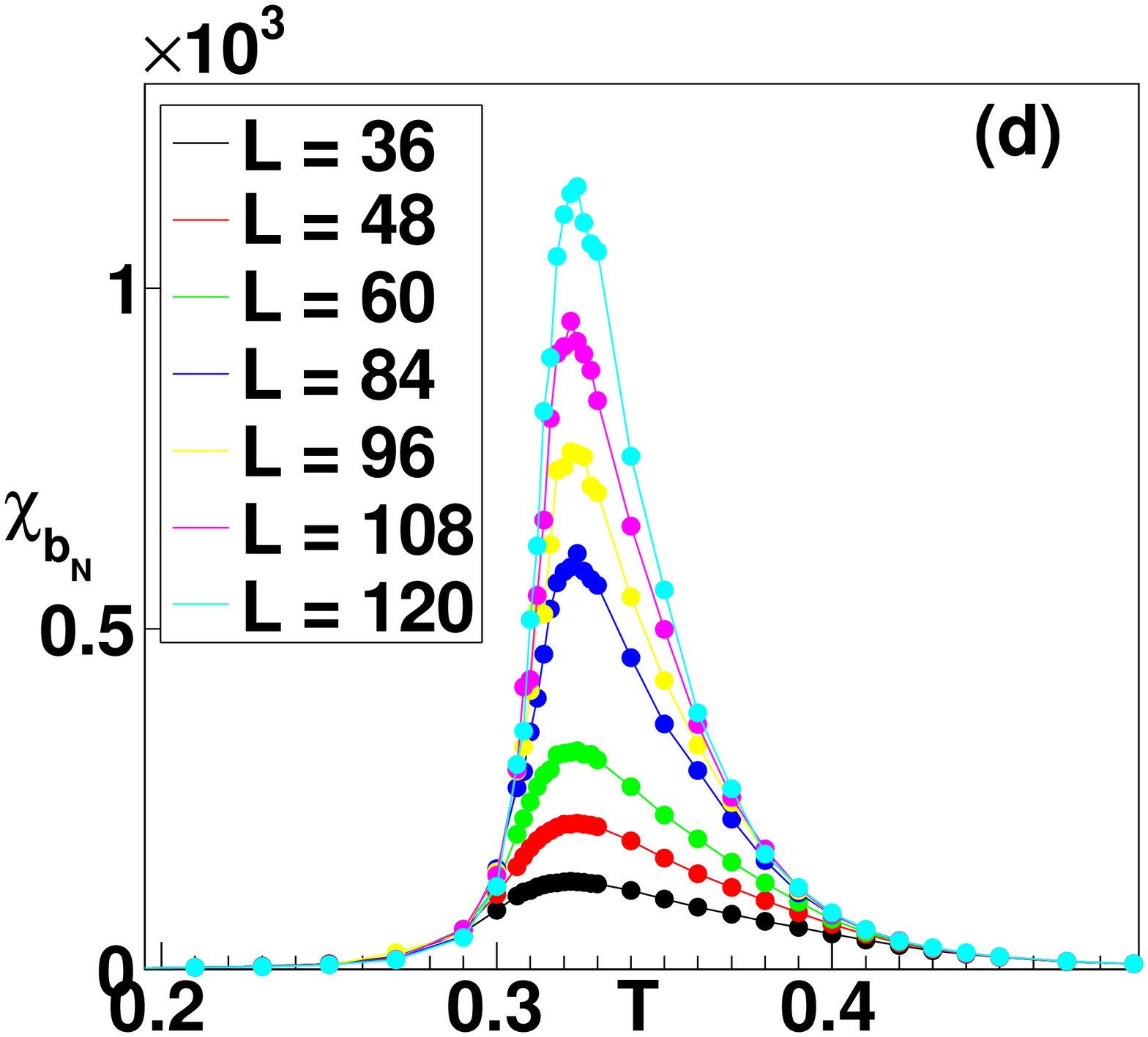}
\includegraphics[width=0.48\columnwidth]{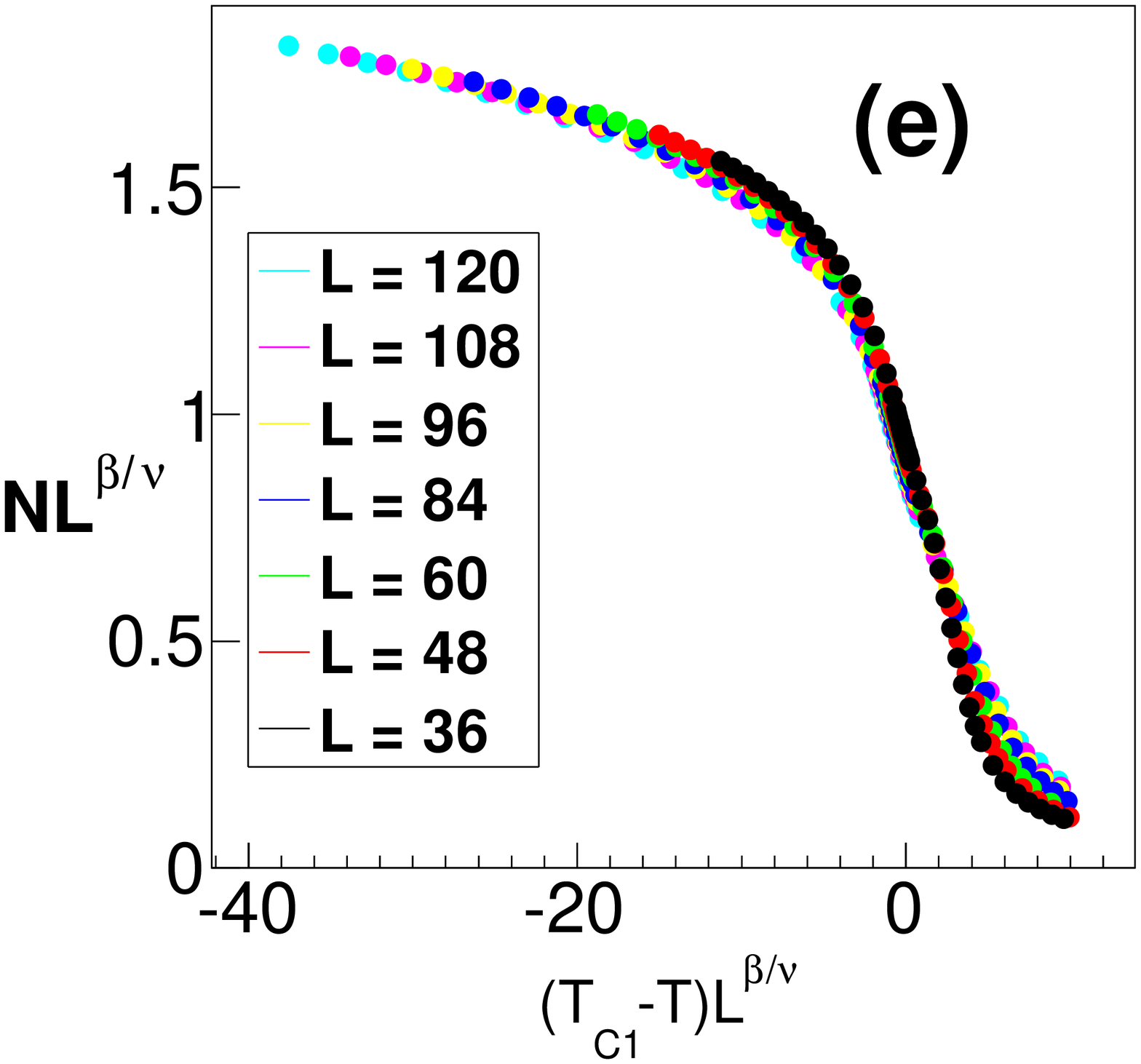}
\includegraphics[width=0.48\columnwidth]{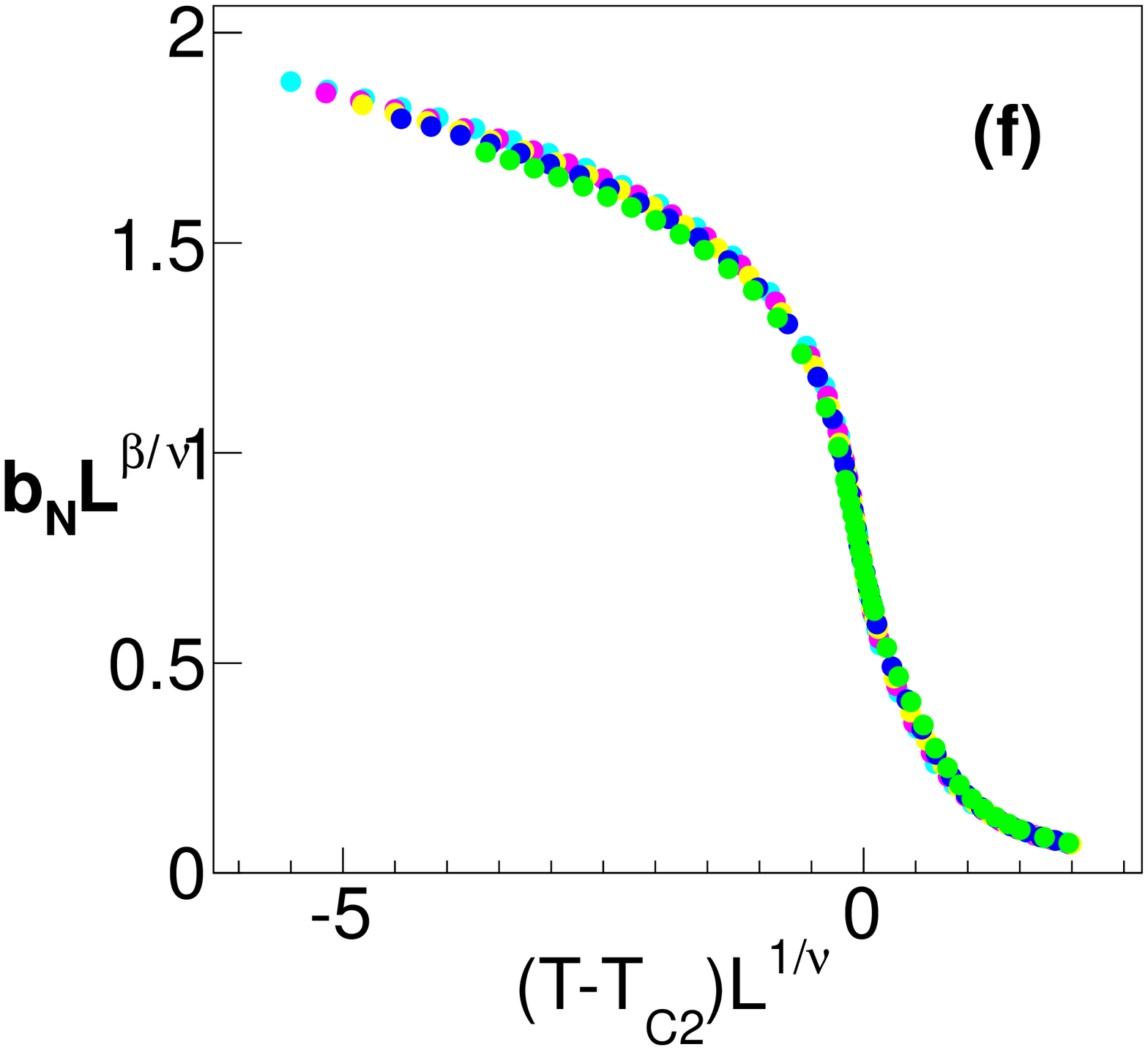}
\includegraphics[width=0.48\columnwidth]{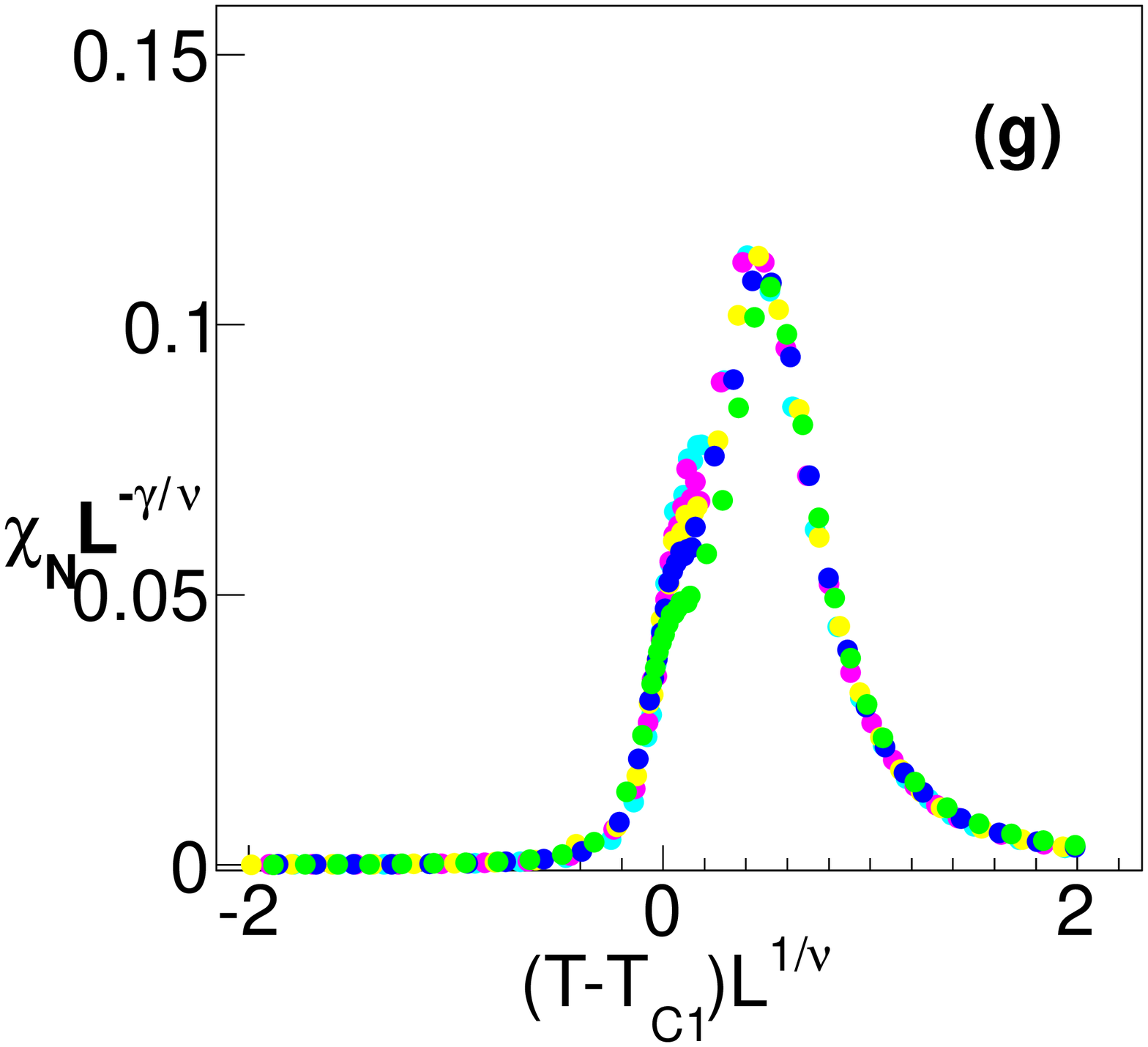}
\includegraphics[width=0.48\columnwidth]{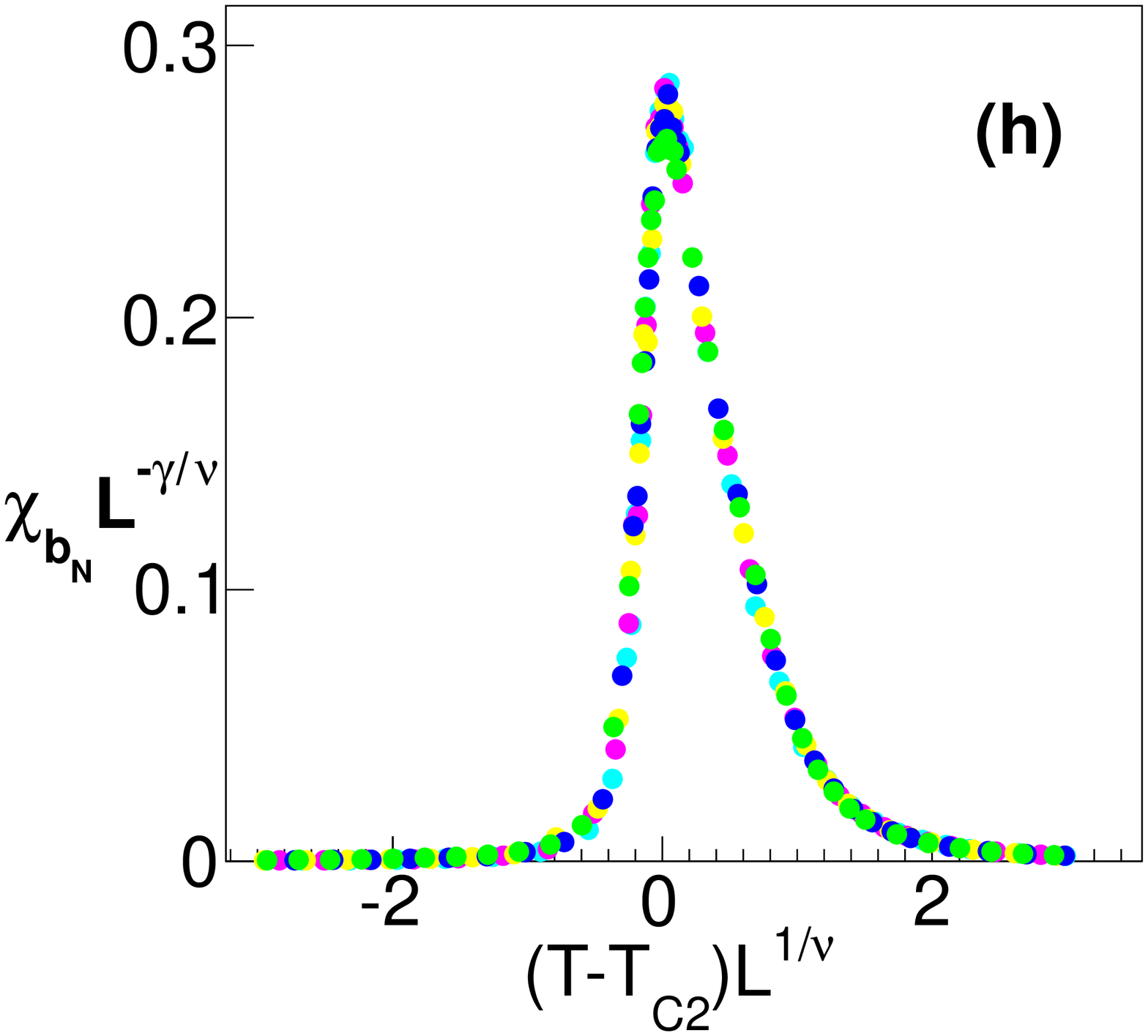}
\caption{The magnetization $N$ (a) and the cubic order parameter $b_N$ (b) as functions of $T$.
The susceptibilities $\chi_{N}$ (c) and $\chi_{b_N}$ (d) as functions of $T$.
(e) and (f): Finite-size scaling of the order parameters $N$ and $b_N$  in the low-temperature region $T<T_{c_1}$.
(h) and (g): Finite-size scaling of the high-temperature susceptibilities $\chi_{N}$  and $\chi_{b_N}$ in the high-temperature region $T>T_{c_2}$.
The anisotropy constant is  considered to be $D=0.1$.}
\label{fig:cubic1}
\end{figure*}

\begin{figure}
\includegraphics[width=0.48\columnwidth]{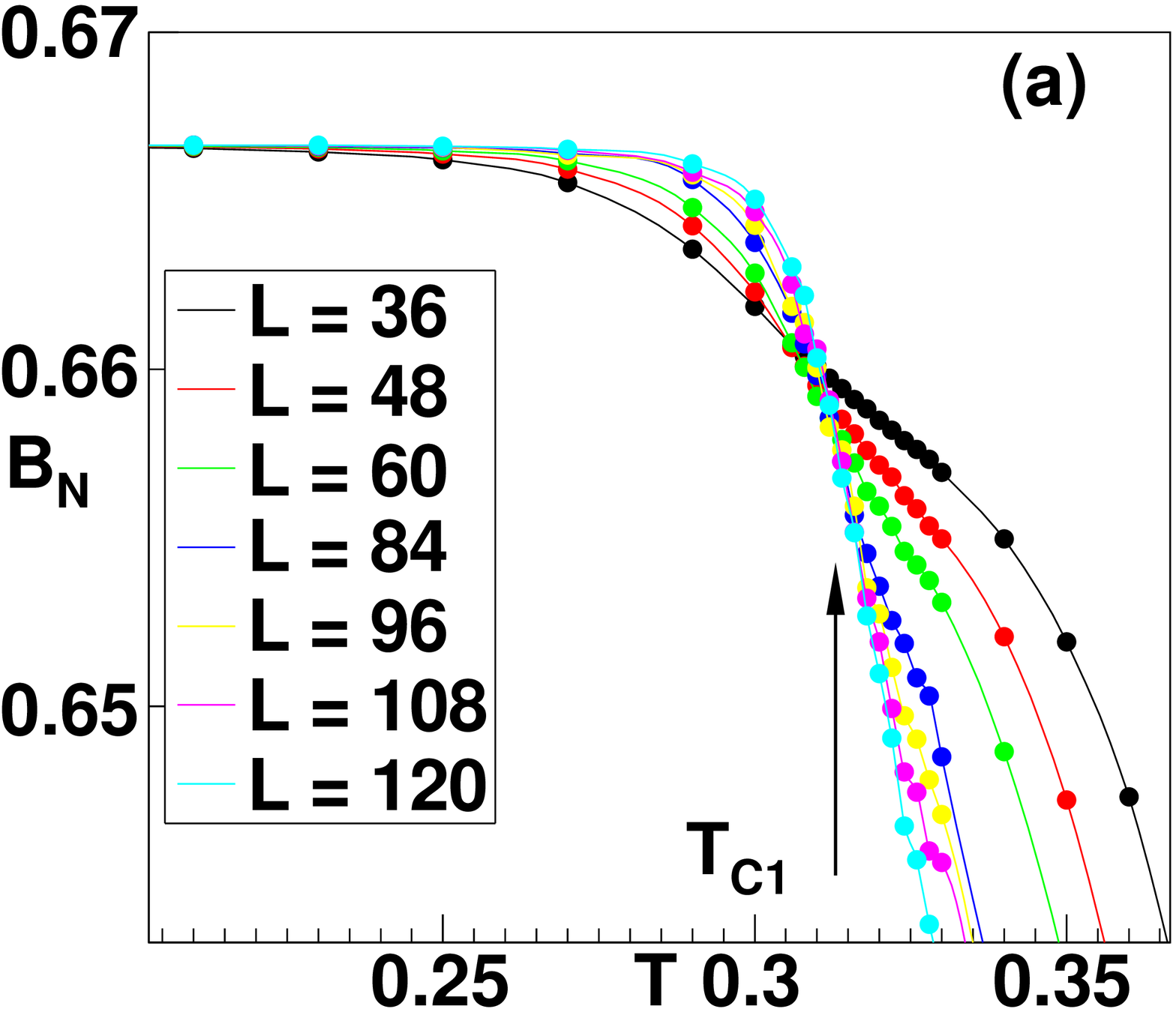}
\includegraphics[width=0.48\columnwidth]{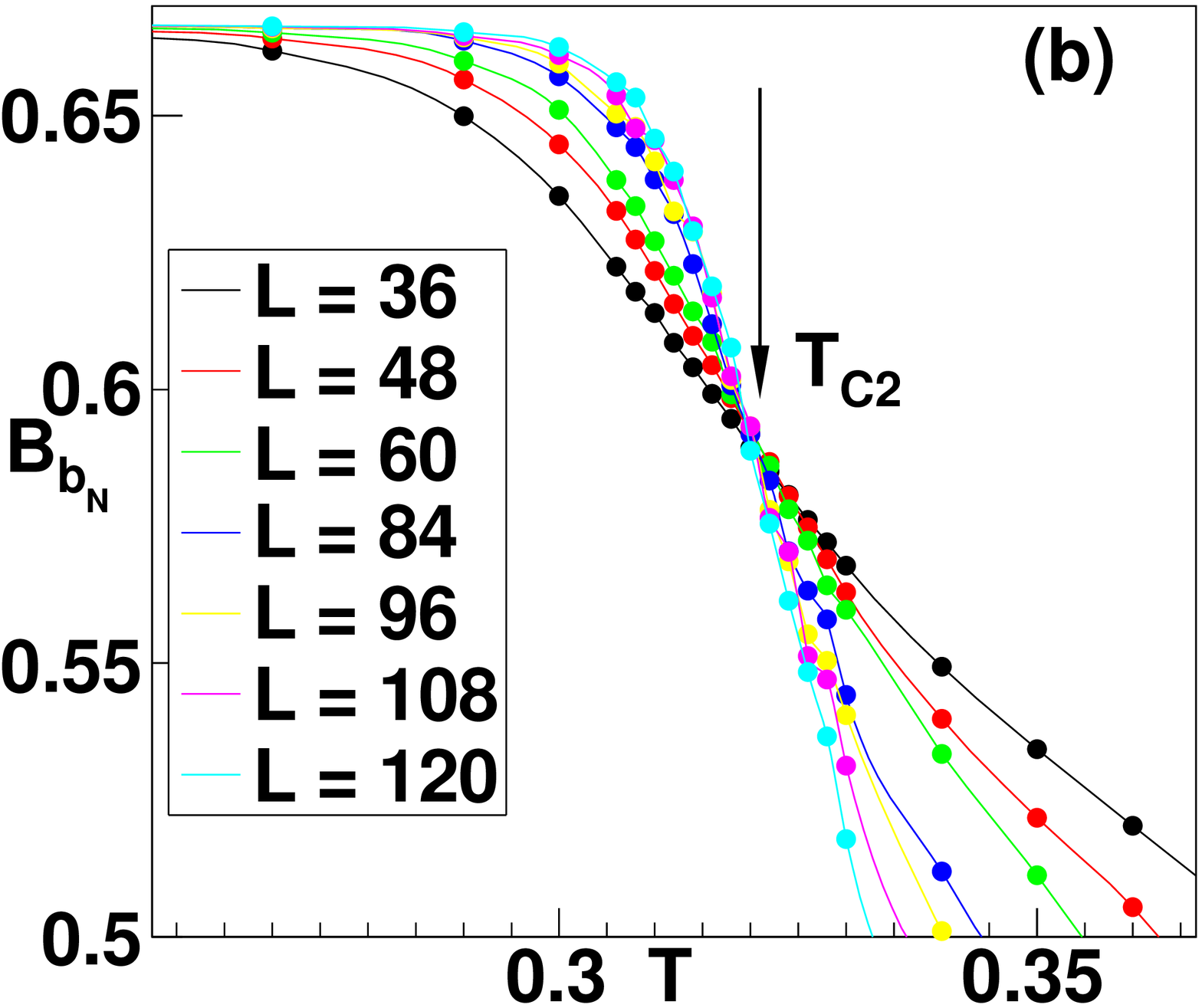}
\includegraphics[width=0.48\columnwidth]{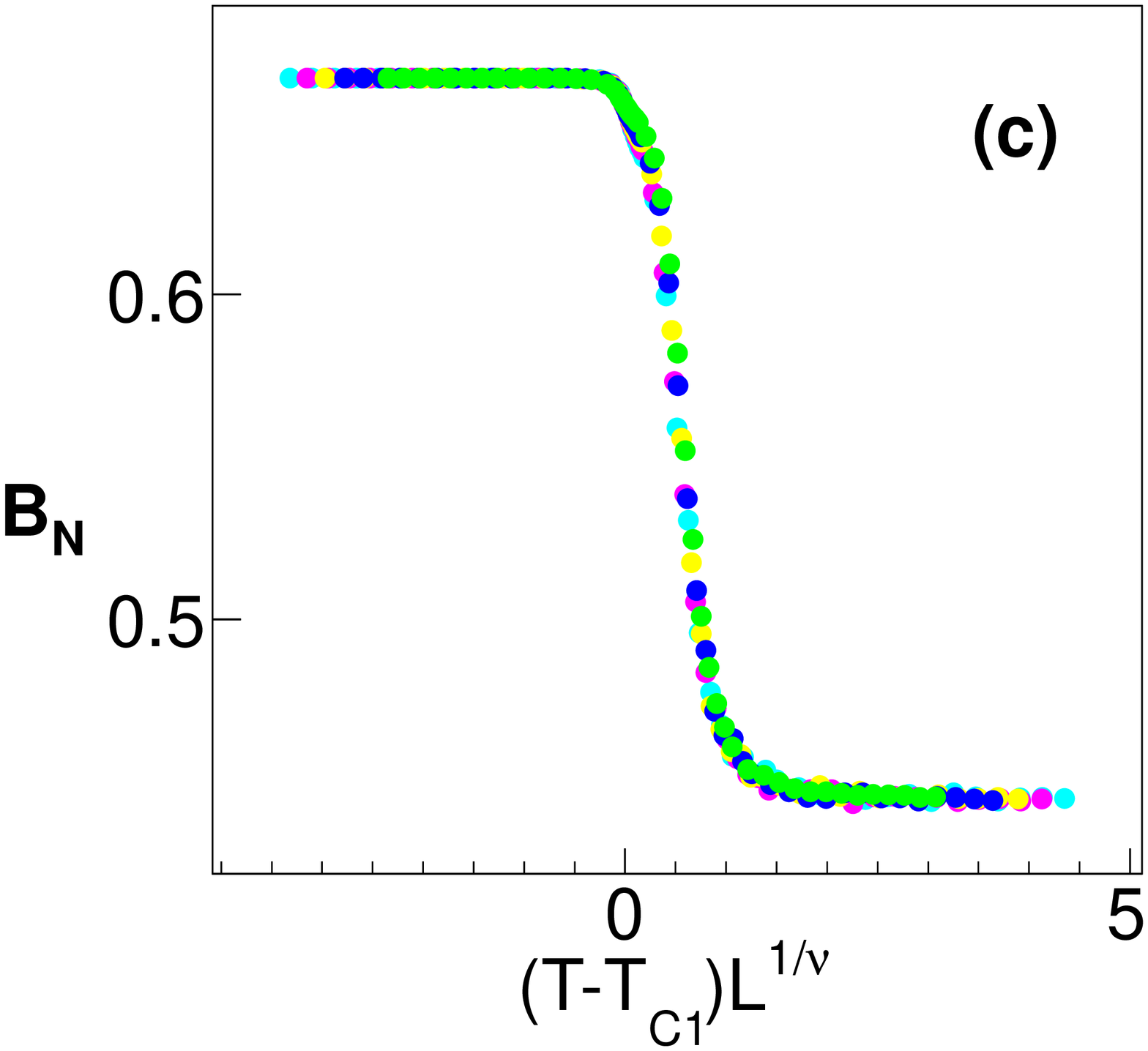}
\includegraphics[width=0.48\columnwidth]{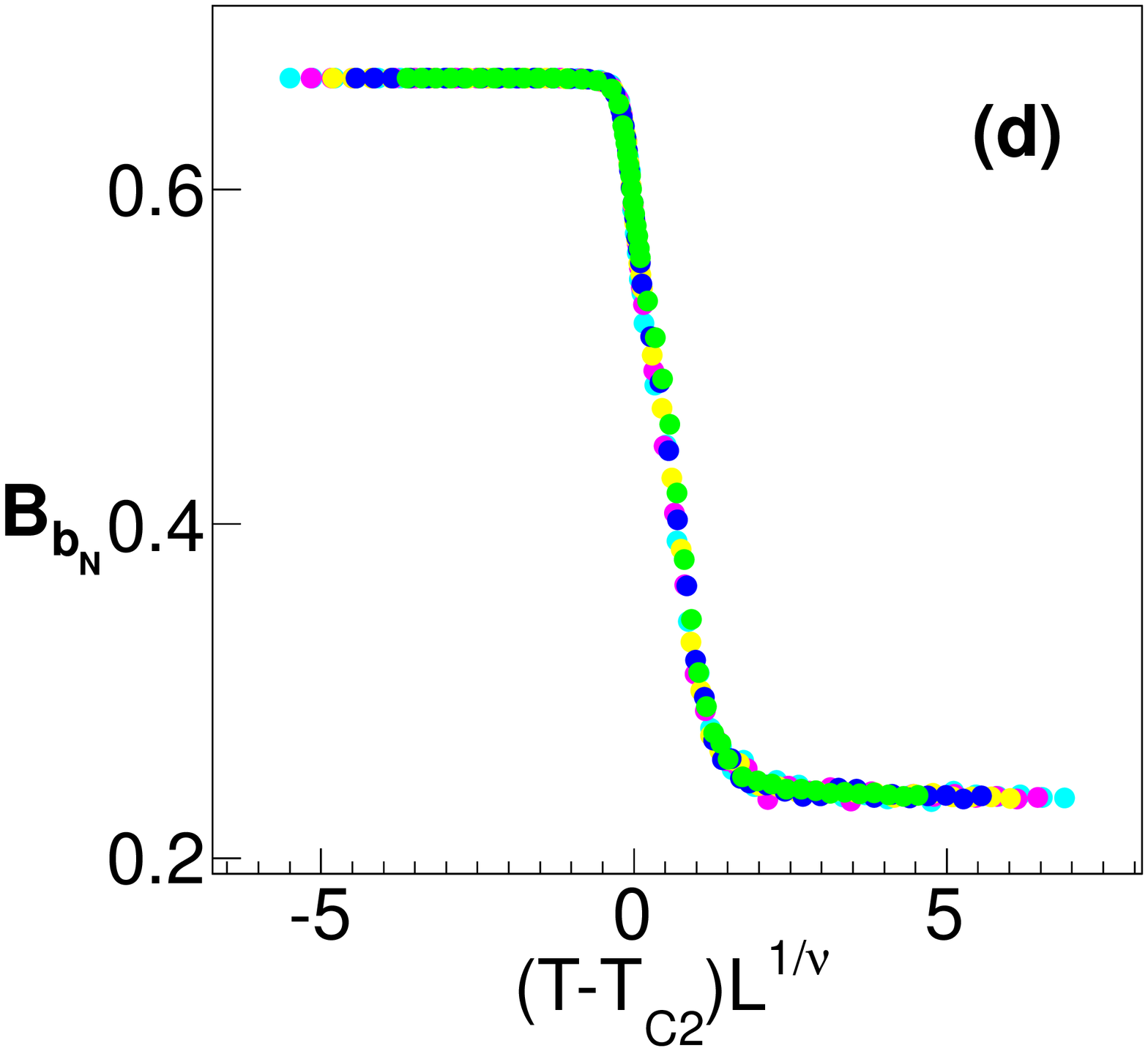}
\caption{
(a) and (b): The Binder cumulants, $B_N=1-\langle N^4\rangle/3\langle N^2\rangle^2$ and $B_{b_N}=1-\langle b_N^4\rangle/3\langle b_N^2\rangle^2$, as functions of $T$.
 The Binder's cumulant crossing points for $N$ (a) and the cubic order parameter $b_N$  give  $T_{c_1} = 0.314\pm 0.001$ and $T_{c_2} = 0.320\pm 0.001$.
(c) and (d): Finite-size scaling of the Binder cumulants $B_N$ and $B_{b_N}$.
The anisotropy constant is  considered to be $D=0.1$.}
\label{fig:cubic2}
\end{figure}

The  numerical results  computed for the value of the cubic anisotropy  $D=0.1$  are presented  in Figs.~\ref{fig:cubic1} (a)-(d).
They display the data for the temperature dependence of the staggered magnetization ${\mathbf N}$, the cubic parameter $b_N$, and corresponding susceptibilities.
In our simulations,  the cubic order parameter $b_N$ is expressed as  the expectation value of a doublet given by
\begin{eqnarray}\label{doublet}
b_{N1} = ({N}_x^2 + {N}_y^2 - 2{N}_z^2)/\sqrt{6} ~,\\
b_{N2} = ({N}_x^2 - {N}_y^2)/\sqrt{2}~,
\end{eqnarray}
where $N_x,\,N_y,\,N_z$ are the components of the N\'{e}el  order parameter $\mathbf{N}$.
At high temperature,  the cubic symmetry is not broken and  $N_x^2=N_y^2=N_z^2$ and $ \langle b_{N1}\rangle  = \langle b_{N_2} \rangle = 0$.
 As we can see from the  Figs.~\ref{fig:cubic1}, both the   order parameters and their susceptibilities  indicate a continuous phase transition at  different temperatures $T_{c_1}$ and $T_{c_2}$ for  $\mathbf {N}$ and $b_{N}$, respectively.
At the higher temperature, $T_{c_2}$, the cubic symmetry is spontaneously broken and  one of the cubic axes is selected, for example, $z$.
Then $\langle b_{N1}\rangle =-\sqrt{2/3}$, $\langle b_{N2}\rangle\simeq 0$  and the cubic parameter $\langle \mathbf{ b_{N}}\rangle $ acquires a finite value.
Nevertheless, there is still no long range order as the time reversal symmetry remains unbroken.
Thus,  we can say that the intermediate phase is nematic-like.
At the lower temperature, $T_{c_1}$,  the time reversal symmetry is also broken and the system acquires long-range magnetic order characterized by non-zero $\langle\mathbf{N}\rangle$.
We can estimate  $T_{c_1}$ and $T_{c_2}$ using the associated Binder's cumulants, $B_N$ and $B_{b_N}$, whose temperature dependencies are presented in Figs.\ref{fig:cubic2} (a)-(b).
 From their crossing points we estimated the transition temperatures to be equal to $T_{c_1} = 0.314$ and $T_{c_2} = 0.32$.

Next, in order to obtain the critical exponents characterizing these two phase transitions we  perform a finite-size scaling  analysis.
As the $\mathbb{Z}_6$  symmetry of the model (\ref{hamcubic})  is reduced to $\mathbb{Z}_2$ symmetry in the intermediate phase, the  high-T transition is expected to be in the universality class of 3-states Potts model and the low-T transition to belong
to the Ising universality class.

First, using the scaling relation for the Binder cumulants, $B_N$ and $B_{b_N}$, we obtained the correlation length exponents $\nu_1$ and $\nu_2$ describing the divergence of the correlation length, $\xi \sim |T-T_{c}|^{\nu(T_c)}$, close to the critical points $T_c=T_{c_1}$ and   $T_c=T_{c_2}$, respectively.
 In accordance with the theory prediction, the best data collapse is obtained for  $\nu_1=0.83$ and $\nu_2=1.0$  which are the critical exponents  of the three-states Potts and 2D Ising model, respectively.

Second, having found  the critical exponents $\nu_1$ and $\nu_2$, we can  estimate  the critical exponents $\beta$ and $\gamma$ by performing a scaling fit of the order parameters and susceptibilities (Figs.~\ref{fig:cubic1} (e)-(g)).
  Near  the critical temperatures, $T_{c_1}$ and $T_{c_2}$, the cubic order parameter and staggered magnetization are expected to satisfy the scaling relation
  ${b_N}\simeq L^{-\beta_1/\nu_1}$ and ${N}\simeq L^{-\beta_2/\nu_2}$, respectively.
  The scaling  laws for their susceptibilities are given by $\chi_{b_N}\simeq L^{\gamma_1/\nu_1}$ and $\chi_{N}\simeq L^{\gamma_2/\nu_2}$.
  The best scaling is obtained for the exponents $\beta_1=1/9$ and $\gamma_1=13/9$  -- the exact scaling coefficients for the Pott's transition, and $\beta_2=1/8$ and $\gamma_2=7/4$ -- the exact scaling coefficients for the Ising transition.

To conclude this section,
in Fig.13 we  present a finite temperature phase diagram for the model (\ref{hamcubic}). We see that the width of the intermediate phase decreases with the increase of the value of the cubic anisotropy parameter $D$.
At around $D\simeq 0.5$, the two transitions become indistinguishable and the intermediate phase collapses.
The transition  between the low-T ordered magnetic phase and high-T disordered phase becomes first order.

\begin{figure}
\includegraphics[width=0.9\columnwidth]{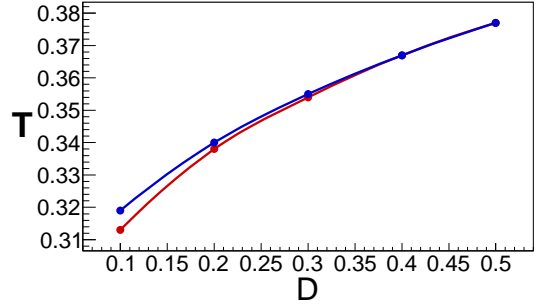}
\caption{Finite temperature phase diagram for the classical honeycomb Heisenberg antiferromagnet   with  the cubic anisotropy (\ref{hamcubic}).
The blue line denotes the high-T three state Pott's transition, and the red line denotes the low-T Ising transition.
The error bars of the calculated quantities are smaller than the size of the circles representing the data points.
}
\end{figure}

\section{Conclusion}\label{conclusion}

In this paper we studied  finite temperature properties of the  classical,  two-dimensional KH  model  and computed the phase diagram of this model in its full parameter space.
We  started by analyzing  the lowest-energy  magnetic configuration and then found that all of the magnetic phases are accompanied by an accidental continuous rotational degeneracy which does not correspond to any symmetry of the Hamiltonian.
 This pseudo degeneracy is lifted by  thermal fluctuations of spins  giving rise to an ordering at low temperature as observed by our MC simulations.
 Specifically we determined that the  low temperature phase is magnetically ordered  at all values  of parameters  for which the model has a discrete symmetry.
 The direction of the order parameter is chosen by thermal fluctuations of the spins through the order by disorder mechanism.

  From numerical MC  simulations, we have  verified that the classical KH model effectively behaves like the six-state clock model.
  At finite temperatures, the model exhibits two  phase transitions and an intermediate phase between them.
  Based on the finite size scaling analysis, we have shown that the intermediate phase is the critical phase with algebraically  decaying correlations of the order parameter.
We  found that  the phase boundaries of the critical phase are of the BKT type.
We also obtained that the numerical values of the critical exponent, $\eta$, characterizing these two transitions are compatible with the theoretical expectations based on the renormalization group analysis for the six-state clock model.\cite{jose77}
 It should be emphasized  that the mapping of the KH model onto the six state clock model is valid only below a certain  temperature which determines the 3D-2D  phase transition, at which the cubic symmetry is already broken.
 The nature of this transition remains to be understood.
  At high temperatures, when $T$ is larger than any energy scale in the system, the effect of thermal fluctuations is to destroy any kind of order and to put the KH model into the  3D paramagnetic state.
    In this regime, the mapping of the classical KH model to the six state-clock model is not valid.

We also performed a comparative study of the honeycomb  Heisenberg  antiferromagnet with  the  cubic anisotropy (\ref{hamcubic}).
  In this case, the continuous symmetry  is  explicitly broken, which allows  the  magnetically ordered phase to  persist up to a finite temperature.
  We have shown that  this model  has distinct finite temperature properties and that  the low-T ordered phase is destroyed in a different way.
  The similarity between the honeycomb  Heisenberg  antiferromagnet with  the  cubic anisotropy, and the  KH model is that the ordered phase is destroyed in two steps.
  The  main differences between the two is that  when the  cubic anisotropy is taken into account explicitly, the intermediate phase is nematic-like and the phase transitions are in the  three-states Potts  and 2D Ising universality classes.

Finally,  we consider the implications of  our  numerical results to  Na$_2$IrO$_3$ and Li$_2$IrO$_3$ compounds.
We have shown that finite temperature magnetic properties of these systems can be captured by the extended KH model.
We found  that for the values of the model parameters relevant to these systems, the model exhibits zigzag magnetic order.
Also, our numerical estimates of the N\'{e}el temperature  are  close to the experimental values.

{\it Acknowledgements.}
The authors are particularly thankful  to  C. Batista, G.-W. Chern, G. Jackeli, Y. Kato, Y. Kamiya and P. W\"{o}lfle for stimulating discussions and many helpful suggestions.
N.P. acknowledges the support from NSF grant DMR-1005932, and also the hospitality of the Aspen Center for Physics and the NSF Grant 1066293 during the work on this paper.


\begin{thebibliography}{100}
\bibitem{interesting} It is interesting to note that emergent U(1) symmetry
of discrete  order parameter has been recently widely discussed in the literature.\cite{isakov03,chern12,wenzel11,sandvik07,orth12}
  For example, an emergent U(1) symmetry is a major prediction of the theory of
the classical $120^{\circ}$ orbital model on the simple cubic lattice,~\cite{wenzel11}
it appears from the non-trivial decoupling of the Heisenberg antiferromagnet on interpenetrating honeycomb
and triangular lattices,~ \cite{orth12}
and is also present  in  the theory of  deconfined critical points for dimer order parameters in the 2D Heisenberg model
with four-spin interactions.~\cite{sandvik07}
\end{thebibliography}
\end{document}